\documentclass[amsmath,floats,floatfix,nofootbib, notitlepage ,hyperref, preprint]{revtex4-1}
\DeclareMathAlphabet{\mathpzc}{OT1}{pzc}{m}{it}
\DeclareMathAlphabet{\mathscrligra}{T1}{calligra}{m}{n}
\usepackage{graphicx}
\usepackage{amsmath}
\usepackage{amsfonts}
\usepackage{calligra}
\usepackage{mathrsfs}
\usepackage{bm}
\usepackage{tensind}
\usepackage{feyn}
\usepackage{PeterStyle}

\begin{document}
\title{Gravitational self-force in non-vacuum spacetimes: an effective field theory derivation}
\author{Peter~Zimmerman\footnote{{\tt pzimmerm@uoguelph.ca}}}
\affiliation{Department of Physics, University of Guelph, Guelph Ontario, N1G 2W1, Canada}
\date{\today}
\begin{abstract}
In this paper we investigate the motion of small compact objects in non-vacuum spacetimes using methods from effective field theory in curved spacetime.  Although a vacuum formulation is sufficient in many astrophysical contexts, there are applications such as the role of the self-force in enforcing cosmic-censorship in the context of the overcharging problem, which necessitate an extension into the non-vacuum regime. The defining feature of the self-force problem in non-vacuum spacetimes is the coupling between gravitational and non-gravitational field perturbations. The formulation of the self-force problem for non-vacuum spacetimes was recently provided in simultaneous papers by Zimmerman and Poisson \cite{Zimmerman:2014uja} and Linz, Friedmann, Wiseman \cite{Linz:2014vja}.  Here we distinguish ourselves by working with the effective action rather than the field equations. The formalism utilizes the multi-index notation developed by Zimmerman and Poisson \cite{Zimmerman:2014uja} to accommodate the coupling between the different fields.  Using dimensional regularization, we arrive at a finite expression for the local self-force expressed in terms of multi-index quantities evaluated in the background spacetime.  We then apply the formalism to compute the coupled gravitational self-force in two explicit cases. 
First, we calculate the self-force on a massive particle possessing scalar charge and moving in a scalarvac spacetime. We then derive an expression for the self-force on an electrically charged, massive particle moving in an electrovac spacetime.  In both cases, the force  is expressed as a sum of local
terms involving tensors defined in the background spacetime and
evaluated at the current position of the particle, as well as tail
integrals that depend on the past history of the particle. 
\end{abstract}
\maketitle 

\section{Introduction}
Solar-mass compact bodies spiralling into super-massive companion objects are promising sources of low frequency gravitational waves for future gravitational wave detectors. The possibility of measuring these signals has fueled research efforts to construct models describing the motion of the small body which go beyond the test-particle approximation. The particle's motion is influenced by the perturbation it sources, causing its orbit to deviate from geodesic motion in the background vacuum   geometry \cite{mino-etal:97, quinn-wald:97}.  The dissipative component of this ``self-force'' acting on the particle causes it  to lose energy and angular momentum to gravitational waves leading to a decay in the orbital radius and an eventual plunge into the larger companion body after a time $T_{\text{inspiral}} \sim M^2/m$.  The gravitational self-force in vacuum spacetimes has developed into maturity over the last decade with notable achievements being: the rigorous formulation in which the compact object is treated as an extended body instead of a point particle \cite{gralla-wald:08,pound:10a}, the numerical computation of the self-force in Schwarzschild spacetime  \cite{Barack:2010tm} and its effect on the innermost stable circular orbit of a Schwarzschild black hole \cite{Barack:2011ed}, the orbital evolution of a massive particle in Schwarzschild spacetime  \cite{warburton-etal:12}, the self-consistent evolution of a scalar particle around a non-rotating black hole \cite{Diener:2011cc}, the computation of the scalar self-force in Kerr spacetime \cite{Warburton:2010eq, Warburton:2011hp}, and the extension to second-order in the mass ratio  \cite{detweiler:12, gralla:12, pound:12a, pound:12b, pound:14}. In addition, the self-force was also successfully compared with numerical relativity \cite{letiec-etal:13} and provided useful information for post-Newtonian fit parameters \cite{shah-friedman-whiting:14}. 

For many applications, the black hole sits in isolation of external fields and matter, in which case the vacuum formulation of the gravitational self-force is adequate. Other applications, however, may require an extension to non-vacuum spacetimes. For instance, one might consider the motion of a satellite around a material body having some companion matter distributed over the region of the spacetime surrounding it.  A non-vacuum formulation will also be required to elucidate the role played by the self-force in scenarios that aim to produce a counter-example to cosmic censorship by overcharging a near-extremal Reissner-Nordstr\"om black hole \cite{poisson-pound-vega:11}.  Hubeny \cite{hubeny:99} has shown that a nearly extremal charged black hole can absorb a particle with parameters living in an open three-parameter family of charge, mass, and energy such that the final configuration's charge-to-mass ratio exceeds the extremal bound leading to a naked singularity. Jacobson and Sotiriou \cite{jacobson-sotiriou:09} showed that this phenomenon is not unique to the Reissner-Nordstr\"om spacetime when they revealed that a near-extremal Kerr black hole can also absorb a particle and be driven beyond the extremal bound. These works treated the particle as a test particle in the black-hole spacetime, and it was soon realized that self-force and radiative effects can play an important role in these overcharging and overspinning scenarios. In fact, Hubeny incorporated approximate self-force effects in her original analysis, Barausse, Cardoso, and Khanna \cite{barausse-cardoso-khanna:10} took into account the gravitational radiation emitted by the particle on its way to overspin a Kerr black hole, and work  by Colleoni and Barack  suggests how the full self-force may be incorporated into an overspinning condition \cite{Colleoni:2015afa}.

Up until recently, the formulation of the gravitational self-force has been restricted to bodies moving in vacuum spacetimes.  In non-vacuum spacetimes, the gravitational perturbation induced by the small body generates a small deviation in the stress-energy tensor of the external field which generates coupled field equations. The coupling between the field equations makes the problem more difficult but not altogether intractable and progress has been made \cite{gralla:10, gralla:13, Zimmerman:2014uja, Linz:2014vja}.   The first full treatment of the coupled self-force on point particles in non-vacuum spacetimes was simultaneously given in Refs. \cite{Zimmerman:2014uja, Linz:2014vja}. In \cite{Zimmerman:2014uja}, regular self-force equations of motion were derived via the Detweiler-Whiting axioms for Lorenz gauge perturbations of both electrovac and scalarvac spacetimes. The treatment in \cite{Zimmerman:2014uja} made use of a condensed index notation which led to  field equations and expressions for the regular and singular solutions for arbitrary external field and point-like source content. The present paper will attempt whenever possible to carry on the same notational conventions as  \cite{Zimmerman:2014uja}. The simultaneous publication by Linz and collaborators \cite{Linz:2014vja} presented the self-force on a charged particle in an electrovac environment. They regularized the fields via a combined approach that used an iterative scheme based on the structure of the perturbed field equations, angular averaging, and mass/charge renormalization. They also derived an explicit expression for the gradient of the singular field and showed that the coupling between the fields does not contribute to the ``singular self-force''; the regularization parameters for the field gradients in the coupled problem are simply the sum of the decoupled parameters.    

The aim of the present work is to provide an additional formulation of the first-order self-force problem for non-vacuum spacetimes using the methods of effective field theory (EFT). The EFT  program is well suited to the binary problem because of its inherent separation of scales. EFT was first successfully applied to the problem of motion for slowly-moving, weakly-gravitating bodies in the post-Newtonian regime \cite{Goldberger:2004jt}, but  it has also proven itself as a contender in the self-force arena through the work of Hu, Galley, and collaborators \cite{galley-hu-lin:06, galley-hu:09, Galley:2010xn, Galley:2011te}. The effective field theory approach to world line dynamics has several attributes including a systematic power counting scheme via Feynman diagrams, dimensional regularization, and internal self-consistency (no ambiguity in defining the world line).  In the EMRI scenario, the EFT approach treats the small compact body as an effective point particle, which takes the form of a skeleton world line with a set of multipole couplings. The effective action preserves the symmetries of the full theory, which in the present case are general covariance, reparameterization invariance, and rotational invariance. This description is sensible at length scales asymptotically greater than the characteristic size of the object, but it breaks down at the body scale just as a far-zone expansion of a radiation field breaks down when approaching the near-zone.   Information about the internal structure of the small object is incorporated by a matching procedure that involves adjusting the coefficients parameterizing finite size effects in the point particle theory to agree with the values obtained using a complete theory describing its internal structure. But, as it has been shown that finite size effects such as  tidal deformations are irrelevant for self-force computations up to fourth order in the mass ratio \cite{galley-hu:09}, such couplings will be ignored.  

 The main results of this work are the equations of motion for a small body due to the coupled self-force in several non-vacuum scenarios.  In all cases we find that the equation of motion can be decomposed into the schematic form \[ m a^\m = F^\m_{0} + F^\m_{\text{local}} + F^\m_{\text{tail}} ,\]
where $  F^\m_{0} $ is the force resulting from the gradient of the background potential, $F^\m_{\text{local}}$ is the local contribution to the self-force which is built from background quantities evaluated on the world line, and  $F^\m_{\text{tail}}$ is the non-local contribution to the self-force which takes the form of a time integral over the particle's past history. 
Here we are not interested in examining the solutions of the field equations. Instead, they are ``integrated out'' from the problem in the process of obtaining an effective action governing only the world line. As a consequence of this, the time-dependent mass of the scalar particle in the scalarvac scenario automatically includes the effects of the regular self-field.  The form of the equation of motion is unaffected by these mass corrections because they are suppressed by an additional power of the small mass ratio. The explicit form of the time-dependent mass due to self-force corrections can be found in  \cite{Zimmerman:2014uja}.  

We begin in Sec. \ref{sec:toy-problem} with a toy problem involving a multi-component scalar field in curved spacetime which is coupled to a world line with $N$ scalar charges. This model captures the general features of the more complicated scalarvac scenario and affords us the opportunity to introduce the EFT procedure for arriving at the effective action via the generating functional. We also introduce several technical tools used in computing the finite self-force such as the
 quasi-local coordinate expansions of various background items in the bulk and around the world line. 
 
In Sec. \ref{sec:mf}, we introduce a general formulation which uses the multi-index notation of  \cite{Zimmerman:2014uja} to derive the effective world line action for a set of coupled perturbations while leaving the field arbitrary. In  Sec.~\ref{sec:causal-eom}, we derive a causal equation of motion from the effective action in terms of retarded Green functions.    Finally, in Secs. \ref{sec:sv}  and \ref{sec:ev}, we apply the formalism to the scalarvac and electrovac spacetimes and derive the coupled self-force equations of motion.  Since its publication, a small error in the local scalarvac self-force was found in the results of \cite{Zimmerman:2014uja}.  The equations of motion are presented here in their corrected form. 
 However, because we do not derive an expression for the singular/regular fields explicitly here, we are unable to do a direct comparison with the results of \cite{Linz:2014vja}. 

\section{Multi-component scalar field}\label{sec:toy-problem}
\subsection{Effective action and formal equation of motion}
Before launching into the complicated coupled problem involving the gravitational field, we choose to examine a simple toy problem which ignores the gravitational field altogether, but which captures the features of the more general case. The model we consider is a multi-component scalar field $\Phi_A(x)$ of dimension $N$ which couples to a massive world line possessing $N$ scalar charges $q_A$. The bulk action for the theory is given by 
\begin{equation}
 S_{\text{bulk}} = -\frac12 \int \left( g^{\a\b} \gamma^{AB}(\Phi) \nabla_\a \Phi_A \nabla_\b \Phi_B + 2 V(\Phi) \right) dV, 
\end{equation}
where $\gamma^{AB}(\Phi)$ is an $N\times N$ dimensional matrix and $V(\Phi)$ is an arbitrary potential even in its argument. The world line action of the theory is a massive point charge interacting with the scalar field according to 
\begin{equation}
S_{\text{pp}} = - m_0 \int d \tau + \sum_{A=1}^{N} q_A \int d \tau \Phi_A.
 \end{equation}
 The motion of the scalar charge in the background spacetime is dictated by 
 \begin{equation}
 m a^\m  =  q^A \left(g^{\mn}+ u^{\m} u^{\n} \right) \nabla_\n \Phi_A,
 \end{equation}
 where we have introduced an effective mass $m := m_0 -   \sum_{A=1}^{N} q_A \Phi_A$ that depends on the scalar field, as well as an Einstein summation convention for ``internal'' indices, $q^A \Phi_A := \sum_{A=1}^N q_A \Phi_A$.
 The presence of the point particle sources a scalar perturbation $f_A$.  The particle will generally also source gravitational perturbations, but we postpone their consideration until later in the paper. The quadratic bulk action is given by
 \begin{align*}
  \delta^2 S_{\text{bulk}}=  - \frac12 \int \Big( & \gamma^{AB}_0 \nabla_\m f_A \nabla^\m f_B +  \nabla_\m \Phi_A \nabla^\m \Phi_B  (\pd^{CD}  \gamma^{AB}_0 ) f_C f_D \\ \no &+  f_C \pd^C \gamma^{AB}_0  \nabla_\m \Phi_A \nabla^\m f_B + V''_0 \gamma^{AB}_0 f_A f_B \Big) dV,
 \end{align*}
 where the ``0'' subscript indicates that the quantity is evaluated on the background scalar field configuration; e.g. $\gamma^{AB}_0 := \gamma^{AB}(\Phi)$.  We further specialize the model by choosing the matrix $\gamma^{AB}_0$ to be independent of the background field  \[ \gamma^{AB}_0 := \mathscr{C}^{AB} = \text{constant} ,\] but retain the freedom that $\mathscr{C}^{AB}$ may contain off-diagonal elements.  The action then reads 
  \begin{equation}
  S_2 =  - \frac12   \mathscr{C}^{AB} \int    \Big(\nabla_\m f_A \nabla^\m f_B  + V''_0  f_A f_B \Big) dV +  \delta S_{\pp} ,
\end{equation}
where \[  \delta S_{\pp} = -  \int \, \, q^A f_A dV .   \]
 By performing an integration by parts we see that the action can be written in standard quadratic form by introducing an operator $O^{AB}$ given by  \[ O^{AB} = - \mathscr{C}^{AB} \left( \Box - V''_0 \right)  ,\] such that 
     \begin{equation}
  S_2 =  - \frac12 \int   f_A O^{AB}  f_B \,  dV  + \delta   S_{\text{pp}}.
\end{equation}
The inverse of the operator $O^{AB}$ defines the propagator of the theory.  The propagator, along with higher $n$-point functions, are obtained  via the moments of the generating functional 
\begin{equation}
	\mathscr{Z}[j] = \int \mathscr{D} f  \, e^{ i S_2} \, e^{ i \int j^A f_A  \, dV},
\end{equation}
where $j^A$ is an arbitrary external source.  The smallness of the particle $m_0 \sim q_A \ll 1$ permits a Taylor series expansion for $e^{i S_{\text{pp}}} $, after which $\mathscr{Z}$ becomes 
\begin{equation}\label{eq:Z-pert1}
	\mathscr{Z}[j] = \int \mathscr{D} f  \, e^{ i  \, \delta^2  S_{\text{\bulk}}} \, \left( 1 + i \delta S_{\text{pp}} + \frac{{i}^2}{2}   \delta S_{\text{pp}}^2 + \cdots \right)  e^{ i \int j^A f_A  \, dV}.
\end{equation}
Replacing $f_A$ by the variational derivative $-i \delta/ \delta j^A$ taken at $j^A=0$, we reduce the above path integral to a Gaussian in curved spacetime.    Although in a general curved background computing the Gaussian path integral is not tractable, it is manageable in flat spacetime, and we may exploit the equivalence principle to design a perturbation series based on local flatness about a point. To do so, we work in a local convex neighbourhood $\mathscr{N}(x_0)$ centred at a \emph{fixed} point $x_0$, and expand deviations from flat spacetime in powers of the coordinate distance from $x_0$.  In this paper we adopt Riemann normal coordinates (RNC) which we denote by $x^\a$. The Riemann normal coordinates of a point $x$ with respect to the fixed base point $x_0$ and tetrad $ e^\a_{\ \m}(x_0)$ are defined by the relation
\begin{equation*}
	x^\a :=  - e^\a_{\ \m}(x_0) \sigma^\m(x,x_0),
\end{equation*}
 where we have introduced Synge's bi-vector $\sigma^\mu(x,x_0)$ which runs tangent to the unique geodesic connecting $x_0$ and a point $x$ in $\mathscr{N}(x_0)$.  The RNC expansion is valid provided that the metric varies slowly on scales of order the size of the convex normal neighbourhood.    A full discussion of bitensors including Synge's function can be found in the review article \cite{poisson-pound-vega:11}. As the field operator ${O}^{AB}$ is built solely from background quantities, it admits the expansion
\begin{align}
{O}^{AB}(x) &= {O}^{AB}(x_0) + \underbrace{{O}^{AB}_{\ \m} (x_0) x^\m +  {O}^{AB}_{\ \m\n} (x_0) x^\m  x^\n  + \cdots}, \no \\
                           &:= \mathscr{F}^{AB}  \quad \, + \quad\quad \quad \quad\quad\quad \mathscr{E}^{AB},
\end{align}
where $\mathscr{F}^{AB}$ denotes the flat spacetime kinetic operator given by
 \begin{equation}
f_A \mathscr{F}^{AB} f_B =  \mathscr{C}^{AB}\,  f_A  \,  \overleftarrow{\pd}_\a \eta^{\a\b} \pd_\b \, f_B,
\end{equation}
which yields the principal part of the field equations upon variation. 
 The quantity $\mathscr{E}^{AB}$ incorporates the curvature corrections to the bulk propagator in the local neighbourhood of $x_0$. For general fields, $\mathscr{E}^{AB}$ will \emph{also} contain contributions from ${O}^{AB}(x_0) $ due the presence of background curvature coupling directly to the perturbed fields, as is the case of vacuum gravity through terms such as $R^{\a\c\b\delta}h_{\a\b}h_{\c\d}$. Such terms persist independently of any expansions the background metric might have. The metric determinant found in the volume element $dV=\sqrt{-g} d^dx$ also contributes curvature terms in the RNC expansion and these must treated with care.  The metric determinant in the volume element of $\exp{ \left ( - \int dV j^A f_A \right)}$ can be eliminated by shifting the external current $j^A \rightarrow \sqrt{-g} j^A$. This shift is of no consequence because we set the current to zero after variation.   To take care of the metric determinant in the bulk action coupling to $O^{AB}(x_0)$, we use the fact that $g$ is expanded around its flat spacetime unit value and we absorb the higher-order terms into a redefinition of $\mathscr{E}^{AB}$. As $\mathscr{E}^{AB}$ is represented by a series in powers of $x^\m$, its contribution to the bulk phase $e^{iS_{\text{bulk}}}$ can be written as a series analogous to that involving the internal source in Eq.~\eqref{eq:Z-pert1} with $f_A$ again being replaced by a variational derivative with respect to the current $j^A$. Explicitly, we find that to second-order in the derivative expansion, the quantity $\mathscr{E}_{AB}$ reads
 \begin{equation}\label{eq:Eab}
f_A \mathscr{E}^{AB} f_B =  f_A \left( C^{AB} \overleftarrow{\pd}_\m \left( \tfrac13 R^{\m \ \n}_{ \ \a \ \b} - \tfrac16 \eta^{\m\n} R_{\a\b} \right) x^\a x^\b \pd_\n +  V''(\Phi) \right)f_A .
 \end{equation} 
 After performing the Gaussian integration 
\begin{equation}
\int \mathscr{D} f \, \exp \left[-\frac{i}{2} \int  f_A  \mathscr{F}^{AB} f_B  \, d v + i \int  j^A f_A \, dv \right] = N e^{-\frac12 j^A \cdot D_{AB'} \cdot j^{B'}} =: Z_0[j],
\end{equation}
where $dv$ is the volume element of Minkowski spacetime and $a\cdot b := \int a(x) b(y)\, dv_x \, dv_y$,  we find 
{ \small
\begin{equation}
\frac{ \mathscr{Z}[j]}{Z_0[j]} = 1 -   \frac12  q^A q^B \iint \, D_{AB}(z,z')  d\tau d\tau' + \frac{i}{2} q^A q^B \iint  d\tau d\tau' \int D_{AC}(z,x) \mathscr{E}^{CD}(x) D_{DB}(x,z')  dv + \cdots 
\end{equation}}
after functional differentiation.
We then Legendre transform  the generating functional with respect to the world line coordinates to obtain a formal expression for the effective action
\begin{equation}
\Gamma[z] = S_{\pp} -  \frac12 q^A q^B \int d\tau \int d\tau' G_{AB}(z,z'),
\end{equation}
where we have used the notation $G_{AB}$ for the complete curved spacetime Green function; i.e., 
\begin{equation}\label{eq:G-expansion}
G_{AB} = D_{AB} - i \int  dv  \, D_{AC} \mathscr{E}^{CD} D_{DB} + \orderof{\mathscr{E}^2} . 
\end{equation} 
Varying with respect to the world line coordinate gives the equation of motion 
\begin{equation}\label{eq:toy-sf}
 m(\tau) a^\a = q^A w^{\a\b} \nabla_\b \Phi_A +  q^A q^B \,\, w^{\a\b} \nabla_\b \int_{-\infty}^{\tau_-} d\tau' G^{ret}_{AB}(z,z'),
 \end{equation}
 where 
 \[w^{\a\b} := g^{\a\b} + u^{\a} u^{\b} \] projects on the spatial plane orthogonal to the world line.  
 The equation of motion must be expressed in terms of the retarded Green function to satisfy causality.
  Although at this stage the retarded propagator does not appear to arise naturally from the effective action, which is time-symmetric under interchange of $\tau$ and $\tau'$, we can ensure that the Green function is retarded by adopting the causal action formalism to break the time symmetry resulting from a naive application of Hamilton's Principle \cite{Galley:2012hx}.  The causal action formulation and the conditions required to obtain the retarded propagator will be discussed in a later section (Sec~\ref{sec:causal-eom}).   
  
 \subsection{Quasi-local expansion}
 As it is written above the equation of motion still suffers from a divergence on the world line due to the singularity present in $G^{ret}_{AB}$ at the simultaneous point $z'=z$. In order to proceed we must regularize the expression, which will allow us to separate the equation of motion into a sum of finite local terms and a tail integral over the particle's past.   The method of choice for regularization is dimensional regularization, which is well adapted to the EFT formulation. Dimensional regularization requires that we work with a momentum space representation for the flat spacetime retarded propagator $D^{ret}_{AB}$.  To regularize the equation of motion,  we must first determine its local behaviour.  Recall that we have a local expansion in RNC for the bulk curvature operator $\mathscr{E}^{AB}$ around a base point $x_0$ as given to second order in Eq.~\eqref{eq:Eab}.   Since $x_0$ is arbitrary we can choose it to coincide with $z^\a(\tau)$, the coordinate of the world line at the ``current time'', so that the origin of the RNC system is comoving with the particle.  Likewise we choose $x^\a$ to be the RNC of $z^\a(\tau')$. What's left is to translate the Riemann coordinates, which are expressed as a series in the geodesic distance between the two points on the world line through Synge's bivector $\sigma_\a$, into the quantities defined on the accelerated world line such as the four velocity and its derivatives. Following \cite{Galley:2010xn}, we translate between quantities written in terms of the geodesic and the accelerated word line connecting the  two points $x$ and $x_0$ by introducing a small parameter $t$ defined as the proper time difference between the current time $\tau$ and some time in the very near past $\tau'$.  Once the quasi-local expansion in terms of $t$ is finished, the expression for the force in terms of a tail integral and an integral over Fourier modes can be manipulated into a form which lends itself to dimensional regularization resulting in a finite expression for the force including local terms. 

We now compute the local force 
\begin{equation}\label{eq:sf-leading-toy}
\mathfrak{f}_\a(\tau) :=  q^A q^B \, w_{\a}^{\ \ \b} \nabla_\b \int d \tau' \,  D^{ret}_{AB}(z,z') , 
\end{equation}  due to the leading-order term in the RNC expansion of the Green function found in Eq. \eqref{eq:G-expansion}, which diagrammatically corresponds to computing graphs without insertions of $\mathscr{E}^{AB}$. 
In Eq.~\eqref{eq:sf-leading-toy}, $w^{a\b}$ is evaluated at the present time $\tau$ and the retarded propagator is expressed in RNC centred at $z(\tau)$ as the Fourier integral 
\begin{equation}\label{eq:prop-flat1}
D^{ret}_{AB}(0,x) = \Lambda_{AB} \, \int_{k,\, \ret} \frac{e^{- i k\cdot x}}{k^2},
\end{equation} where $\Lambda_{AB}$ is defined by the relation $\Lambda_{AC} \Box^{-1} \mathscr{F}^{C}_{\ B} = \delta_{AB}$, $k^2$ is the Minkowski space inner-product of wave-vectors $k^2 = \eta_{\m\n} k^{\m} k^{\n} = -\omega^2 + \vec{k}^2$, and the subscripted integral corresponds to a $d$-dimensional Fourier integral using the retarded contour in the $\omega$-plane. The $\omega$ integral has  two poles sitting at plus and minus the magnitude of the spatial vector $\vec{k}$, which forces us to choose a contour for computing the residues. The subscript $\ret$ indicates that we have chosen a contour that passes over the poles and closes at $\text{Im} \, \omega \rightarrow - \infty$  corresponding to the condition $\tau > \tau'$.  Then, using the definition of RNC we make the substitution $x^\a(z') = -e^\a_{ \ \b}(z) \sigma^\b(z,z') $, and orient the tetrad such that $e^{\a}_{\ 0} = u^{\a}$. This allows the exponential in $D^{ret}_{AB}$ to be expressed in terms of $\sigma_\a(z,z')$ as defined on the geodesic connecting $z$ and $z'$ as $e^{i k \cdot \sigma}$.  In terms of the accelerated world line, the geodesic bivector has the expansion
\begin{equation}\label{eq:sigma-expansion}
	\sigma_\a(z,z') = - t u_\a - \frac{t^2}{2!} a_\a - \frac{t^3}{3!} {\dot a}_a + \cdots,
\end{equation}
where $a_\a$ is the covariant acceleration $a_\a = \dot{u}_\a := D u_\a /d \tau$.  Using the above expansion for Synge's bivector, along with the definition of RNC, we find that the exponential in the Green function takes the expanded form 
\begin{equation}\label{eq:exp-exansion}
	e^{ i k \cdot x} = e^{ i k^{0} t  } \left( 1 - i k_\a \sum_{n=1}^\infty \frac{t^{n+1}}{(n+1)!} \frac{D^{n}}{d \tau^n} u^\a + \cdots \right),
\end{equation}
where the numbered/latin indices indicate frame components; e.g., $k^\a = k^{ 0} e^\a_{ { 0}} + k^{ { i} } e^\a_{\ { i}}$. Changing integration variables from $\tau'$ to $t$ using $t=\tau'-\tau$, we find that all integrations necessary for the local expansion of the force involve evaluating a single master integral 
\begin{align}\label{eq:I-master}
	I _{i_1 \ldots i_q}(m,n,p,q) &:= \int_{-\infty}^{\infty} dt \int_{\vec{k}} \int_{k^{ 0}} \frac{ e^{ i k^{ 0} t} }{ ( \vec{k}^2 - (k^{ 0})^2 )^m } t^n (k^{ 0})^p k_{{i}_1} \cdots k_{{i}_q} , \no \\
	&= \frac{ i^{2n-p} }{ 2^{d-1} \pi^{\frac{d-1}{2}} } \frac{ \Gamma\left(\frac{d+q-1}{2}\right) \Gamma\left(m-\frac{d+q-1}{2} \right) }{ \Gamma\left(\frac{d-1}{2}\right)\Gamma(m) }\frac{ \left(d+p+q-2m-1\right)! }{ \left(q+1\right)!! } \delta_{i_1 \ldots i_q} \delta_{n,d+q+p-2m-1}
\end{align}
where ${ \delta}_{i_1 \ldots i_q}$ is a sum  over the product of Kronecker deltas with all possible permutations of $q$ indices, ${ \delta}_{i_1 \ldots i_q} = \delta_{{ i}_1 { i}_2} \cdots\delta_{{ i}_{q-1} { i}_q} + \text{perms}$. The above integral is an extension of the master integral used in \cite{Galley:2009px}, which was previously specialized to the single $m=2$ case.   The integral is straightforwardly performed after making a few observations about its form.
The first is that odd powers of spatial momenta yield zero by rotational invariance. Secondly, powers of $t$ can be written as derivatives with respect to $k^0$ after integrating by parts. After integration by parts, the $t$ integral becomes a delta function $\delta(k^0)$, and the spatial $\vec{k}$ integral is  reduced to a standard form that can be performed via the Beta function
\begin{equation*}
\int_0^{\infty} \frac{ k^{b-1} }{(k^2 + v^2)^m} dk = v^{b-2m} \frac{ \Gamma(b/2) \Gamma(m-b/2) }{ 2 \Gamma(m) }.
\end{equation*}
  The result of the Euclidean integral, along with the distribution $\delta(k^0)$ coming from the $t$ integral,  results in the Kronecker delta $\delta_{n,d+q+p-2m-1}$ whose arguments are determined by the exponent of  $k^0$ which must be zero otherwise the whole expression is zero  \cite{Galley:2009px}.   The upshot of this is that $t$ power counts like $1/ { \abs{ \vec{k}}} $ and $k^0$ doesn't need to be power counted when determining the high energy behaviour of the integral.

Using the above expansions in conjunction with the master integral we find just a single non-zero term in the series \eqref{eq:exp-exansion} involving the integral $I_{ij}(1,3,0,2) =
\frac{1}{2\pi} \delta_{ij} $ leading to the local self-force expression
\begin{equation*}
 \mathfrak{f}_\a = \frac{1}{4\pi} \frac{1}{3} \Lambda_{AB} q^A q^B \, w_{\a\b} \dot{a}^\b .
\end{equation*}
Notice that in the limit $A=B=1$, the local self-force above reduces to the expected result for a single charge \cite{Quinn:2000wa}.

Let us now examine the effect of curvature terms in the bulk action on the local self-force. The presence of bulk curvature entering through the quantity $\mathscr{E}_{AB}$ causes a focusing of null rays and the detailed history dependent force which results is generally so complicated that it must be evaluated numerically or left as a formal expression.
 The local nature of the force in curved spacetime, however, can be established in the normal neighbourhood of the particle where unique geodesics exist. The starting point for the computation is the RNC representation of $\mathscr{E}^{AB}$  given in Eq.~\eqref{eq:Eab}.   Power counting, we find that the high energy behaviour of the potential term  in four dimensions  is $\orderof{\lambda^{-2}}$  for a cutoff $\lambda \sim (\text{size of body})^{-1}$ and will not lead to a local term.  To second-order in the RNC expansion we find that the propagator \eqref{eq:G-expansion} reads
\begin{equation}\label{eq:G-toy}
G_{AB}(x_0,x) = D_{AB}(x_0,x) - \frac13 \mathscr{C}^{CD}  \left( \delta^{ ( \m }_\a \delta^{ \n )}_\b - \frac12 \eta^{\mn} \eta_{\a\b} \right) R^{\a \ \b}_{\ \c \ \d} \int_y y^\c y^\d \pd_\m D_{AC}(x_0,y) \pd_\n D_{DB} (y,x) + \cdots
 \end{equation}
 where the Riemann tensor is evaluated at the base point $x_0$ at which the Green function $D_{AB}$ has the form \eqref{eq:prop-flat1}. We find that the term $   R^{ (\a \ \b)}_{\ \c \ \d} \int_y y^\c y^\d \pd_\a D_{AC} \pd_\b D_{DB} (y,x')$ vanishes algebraically and the only contributing part at second-order in background derivatives reads 
 \begin{equation}
 \frac16 \mathscr{C}^{CD} R_{\a\b} \int_y y^\a y^\b \pd_\m D_{AC} \pd^\m D_{DB}  = - \frac13 \Lambda_{AB} \int_k  \frac{ e^{i k \cdot x}}{k^6} \left( R k^2 - 2 R_{\a\b} k^\a k^\b \right)  ,
 \end{equation}
 where we have eliminated $y^\a$ using the relation $y^\a e^{i k \cdot y} = -i \frac{\pd e^{i k \cdot y} } {\pd k_\a}$. After inserting the expression above into Eq~\eqref{eq:toy-sf}, we find the local self-force containing  curvature terms is given by
  $ \frac16  \, \Lambda_{AB} q^A q^B \, w_{\a\b} R^{\b\c} u_\c$ as expected from the known result of the self-force on a single charge.
 \subsection{Equation of motion}
Assembling the local expressions,  we find that the explicit self-force equation of motion reads 
\begin{equation}
 m(\tau) a^\a_{\text{self}} =  \frac{1}{4\pi}   \, \Lambda_{AB} q^A q^B \,  w^{\a}_{ \ \b} \left(    \frac{1}{3} \dot{a}^\b +  \frac16  \, w_{\a\b} R^{\b\c} u_\c \right) +  q^A q^B \,\, w^{\a\b} \nabla_\b \int_{-\infty}^{\tau_-} d\tau' G^{ret}_{AB}(z,z').
 \end{equation}

\section{Multifield formalism}\label{sec:mf}
We now consider a more general scenario in which a small non-rotating but otherwise arbitrary object sources a set of perturbations of a background non-vacuum spacetime. The background spacetime, 
$\mathscr{M}_g$, is defined by a smooth metric satisfying the non-vacuum Einstein field equations in the absence of the small body (or a suitable limit where the mass, charge, and size of the body go to zero uniformly).
We characterize the small body by a set of 
internal sources, which we collectively denote by a set of local quantities $\mu^A$, characterizing the multipoles of the body. In the multi-scalar field toy model, the source $\mu^A$ took the form $\mu^A = - q^A \int d\tau \delta(x,z)$.  The source generates a perturbation of the background fields, which we denote collectively
by the field multiplet $\psi_A$. Eventually, we will consider the set of sources $\mu^A = \{\rho,\, j^\a,\, t^{\a\b} \}$, where $\rho$ is the scalar charge density, $j^\a$ is the electric current, and $t^{\a\b}$ is the stress energy of the world line, along with the corresponding set of field perturbations
 $\psi_A = \{ f,\, b_{\a},\, h_{\a\b} \} $.  With this notation, the discussion very closely parallels the treatment of the multi-scalar field toy model. 
 
We begin by constructing  a bulk action quadratic in the fields which we add to the phenomenological action characterizing the extended body as a point particle with a set of multipole couplings. We construct the quadratic action by applying a perturbation operator to the 
full non-linear Lagrangian of the theory and integrating the result over the background manifold. For our purposes we define the $n$-th order perturbation of a quantity $Q$ as $\delta^{n}\, Q := \frac{1}{n!}\frac{d^n Q}{d\epsilon^n} \vert_{\epsilon=0}$. In a local coordinate chart of the background manifold with orientation $\epsilon_{0123}=1$ having four-volume weight $\sqrt{-g}$, the second-order perturbation of the action reads
\begin{equation}\label{eq:second-order-action}
	S_2 = - \frac{1}{2} \int \psi_A O^{AB} \psi_B\, dV - \int \, \mu^A \psi_A dV =: \delta^2 S_{\text{bulk}} + \delta S_{\text{pp}},
\end{equation}
where $dV$ is the invariant volume of the coordinate chart. As we did with the toy-model, we separate $O^{AB}$ into a part $\mathscr{F}^{AB}$ which is invertible in flat spacetime, and a curvature operator $\mathscr{E}^{AB}$ which we treat perturbatively in the local neighbourhood of the body.  
In writing Eq.~\eqref{eq:second-order-action} we've assumed that the field variations (and their first derivatives) vanish sufficiently fast at large distances to ensure boundary terms can be discarded when performing integration by parts.
 
 As we did with the multi-scalar toy model, we start constructing the  effective action by introducing  a generating functional which produces all  the possible field correlations  of the theory. To build the generating functional we introduce external sources $J^A$ which couple linearly to the fields. The generating functional as a functional of the external current coupled to the field reads 
\begin{equation}
	\mathscr{Z}[J] = \int \mathscr{D} \psi  \, e^{ i S_2} \, e^{ i \int J^A \psi_A  \, dV}.
\end{equation}
As $\mu^A$ is perturbatively small, we express $e^{i \delta S_{\text{pp}}} $ as a series expansion, after which $\mathscr{Z}$ becomes 
\begin{equation}\label{eq:Z-pert2}
\mathscr{Z}[J] = \int \mathscr{D} \psi  \, e^{ i \, \delta^2 S_{\text{\bulk}}} \, \left( 1 - i \int \mu^A \psi_A dV - \frac12   \iint \, \mu^A  \mu^{B'} \psi_A  \psi_{B'}  \, dV dV' + \cdots \right)  e^{ i \int J^A \psi_A  \, dV},
\end{equation}
just as for the toy-model. The right-hand side of Eq.~\eqref{eq:Z-pert2} is simplified by replacing $\psi_A$ by the variational derivative $-i \delta/ \delta J^A$ evaluated at $J^A=0$.  With all powers of $\psi$ removed we are left with a Gaussian integral in curved spacetime. In the toy model we simply evaluated the Gaussian and proceeded from there. However, performing the Gaussian integral requires $\mathscr{F}^{AB}$ to be invertible. If $\psi$ contains fields with gauge symmetries,  the inversion is problematic because $O^{AB}$ becomes a singular matrix and the path integral diverges due to over counting physically equivalent field configurations.   To remedy this, we adopt the Faddeev-Popov prescription which involves adding an additional quadratic contribution to $S_2$ given by
\begin{equation}\label{eq:S-gf}
S_{\text{gf}} = - \int \, G_A G^A \, dV 
\end{equation}
where  $G^A$ is a gauge-fixing constraint acting on the fields.  We state without proof that the Faddeev-Popov prescription gives a satisfactory result without introducing spurious gauge modes. Further details regarding the Fadeev-Popov prescription will be elaborated in the appendix. We absorb the extra terms resulting from the gauge fixing procedure, which  involves several integrations by parts, into a redefinition of $O^{AB}$, which we will call 
$\tilde{O}^{AB}$. The gauge-fixed operator $\tilde{O}^{AB}$ is  invertible (at least in a perturbative sense).  

Following the steps in the toy model, we again work in the normal neighbourhood of a point and write the inverse Green function $\tilde{O}^{AB}$ as a coordinate expansion where the leading order term $\mathscr{F}^{AB}$ is the invertible in flat spacetime and curvature corrections are contained in the quantity  $\mathscr{E}^{AB}$.   With this, the curved spacetime propagator takes the form of \eqref{eq:G-expansion} as it enters the generating functional 
{ \small
\begin{equation}
\mathscr{Z}[J] =  \left( 1 - \frac12 \iint \mu^A \, G_{AB'} \mu^{B'} \, dV \, dV'  \right)Z_0[J].
\end{equation}}

So far we have left the internal source $\mu^A$ arbitrary, merely requiring that it be small in magnitude relative to the scale set by the background. To proceed further we consider the specific case for which $\mu^A$ is an effective point-particle world line.  For this description to make sense in the context of effective field theory, we should also add additional higher-order couplings involving non-minimal ``operators''  which encode information about the internal structure of the body. However, it has been shown that these higher order terms are irrelevant for a first order self-force calculation, and so we ignore such terms here \cite{galley-hu:09}.  A point source amounts to the choice of currents
\begin{equation*}
	\mu^A(x) = - \int d\tau' \, \delta(x,z') s^A(z'),
\end{equation*}
where $s^A$ specifies the type of current in question. Since we have now introduced a point particle degree of freedom into the theory, we must also introduce an external source $\zeta^\a$ coupling to it, and from it we build a generating functional for the entire system
\begin{align}\label{eq:gen-func-pp}
	Z[J,\zeta] &= \int \mathscr{D}z \,\, e^{i \int \!\zeta^\a z_\a\, dV}   \mathscr{Z}_{\text{pp}}[J],
\end{align}  
where the subscript pp serves to indicate that the point particle source is to be used for $\mu^A$.  Although we have explicitly introduced a coordinate dependence into the generating functional, this coordinate dependence will ultimately cancel in the effective action. The connected portion of the generating functional is obtained via a logarithm.  Performing a Legendre transform of the connected part with respect to the external current coupled to the world line yields the effective action
\begin{equation}
\Gamma[z] = - i \, \ln Z - \int \zeta^\a z_\a dV,
\end{equation}
which generates all one-particle irreducible connected diagrams.  Quantum fluctuations of the world line are so highly suppressed for macroscopic astrophysical bodies that we may treat $z$ as a classical word line. The upshot of this is that the integration over all world line configurations is easily performed using the saddle-point approximation.  With the addition of the zeroth order point-particle action $S_{\text{pp}}^0$, which is responsible for accelerated motion in the background, we arrive at the effective action for our problem
 \begin{align}\label{eq:eff-action}
 	\Gamma[z] &= S_{\text{pp}}^0 - \frac12 \iint d\tau d\tau' \, s^A D_{AB'} s^{B'} + \frac{i}{2} \iint d\tau d\tau' \, \int d v\,\,  s^A D_{AC} \mathscr{E}^{CD} D_{DB'} s^{B'} + \cdots \\ \no
	& := S_{\text{pp}}^0 - \frac{1}{2}  \iint  d\tau d\tau' \, s^A G_{AB'} s^{B'}  ,
\end{align}
	where $\mathscr{E}^{CD}$ is evaluated at a bulk point in Riemann normal coordinates. Although the series for $G_{AB}$  proliferates indefinitely, on the basis of dimensional analysis only the first two terms which we have explicitly written down contribute a local self-force. All higher order terms ($\orderof{\mathscr{E}^2}$ and above) are contributions to the tail which we will eventually represent in terms of a formal expression.
\section{Causal equations of motion}\label{sec:causal-eom}
To derive an expression for the self-force using an action principle, one must adopt a formalism where the effective action describes causal evolution under the effects of radiation reaction. 
In the quantum field theory of non-equilibrium systems \cite{Schwinger} this formalism is commonly referred to as the IN-IN construction or closed-time-path (CTP) formalism. It differs from the more common IN-OUT formalism, where scattering boundary conditions are placed at the remote future and remote past and the propagator of excitations is the Feynman propagator which is symmetric on interchange of spacetime event points. In the IN-IN formalism the IN states are prescribed and the future states are determined by the evolution of the IN states under the action of the generator of time translations. A classical formulation of this has been introduced by Galley under the moniker Causal Action Principle (CAP) \cite{Galley:2012hx,Galley:2014wla}. Under the CAP construction one breaks the time reversal symmetry by introducing two copies of each configuration variable in the action. 
One then imposes the conditions that both variables are stationary at some initial time (zero variation) and that at some final time, which may be in the remote future,  the two configurations are equal but their values are left unspecified.  Causal equations of motion are obtained by varying  the paths and then setting them equal to each other. 

The CAP action for this theory is 
\begin{equation}\label{eq:CPT-action-def}
 S = S_1 - S_2 := S_{\bulk}[\psi_{A,\,1}]- S_{\bulk}[\psi_{A,\,2}] +  S_{\text{pp}} [z^\a_1, \psi_{A,\,1}] - S_{\text{pp}}[z^\a_2, \psi_{A,\,2} ],
\end{equation}
where $S_{1}$ is the action obtained by evaluating all quantities on the ``1'' configuration and $S_{2}$ is the action obtained by evaluating all quantities on the ``2'' configuration. 
The 1-2 index structure of the RHS above suggests the introduction of a 2-d space with  CAP indices $a,b,c,..=1,2$  with Lorentzian metric
\begin{equation}
 C_{ab} = C^{ba} = \text{diag}(1,-1).
\end{equation}

To produce the generating functional we couple the field configurations and world lines to external currents. With these considerations, the generating functional for IN-IN correlation functions in this theory takes the form
\begin{align}\label{eq:in-in-gen-func}
Z[J,\zeta] = &\int_{\CTP} \mathscr{D}z \, \mathscr{D}\psi \, \exp \Bigg[ iS + i C^{ab}\left( \int  \zeta_a^\a z_{b\a}  + \int  J_a^A \psi_{bA}    \right)\Bigg].
\end{align}
The functional measure for the path integral is defined by the conditions that the configurations are equal to some unspecified value at a time $t=t_f$ in the remote future and that their values are specified on a spatial slice $\Sigma(t_i)$ in the remote past by suitable initial data
\begin{equation}
 \int_{\CTP} \mathscr{D} \psi  = \int_{\Sigma(t_f)} \mathscr{D} \psi(t_f,\vec{x}) \int_{\psi_1(t_i,\vec{x})}^{\psi(t_f,\vec{x})}  \mathscr{D} \psi_{1} \int_{\psi_2(t_i,\vec{x})}^{\psi(t_f,\vec{x})}  \mathscr{D} \psi_{2},
\end{equation}
As we are only interested in classical trajectories of the particle, the measure for the particle's paths can be ignored.
The propagators for the theory are given by functional differentiation with respect to the external currents
\begin{align}
	D^{ab}_{AB'}(x,x')  &= - \frac{1}{Z} \frac{ \delta^2 Z }{ \delta J_a^{A}(x) \delta J_b^{B'}(x')}\Bigg|_{J=0}. \\ \no
\end{align}
Notice that the two-point function is now a matrix-valued quantity, each element of which corresponds to propagation on a different branch of the CTP.  The matrix $D^{ab}$ for a scalar field has the form 
\begin{equation}
  D^ {ab}(x_0,x) = \left(
\begin{array}{cc}  D_F(x_0,x) & - D_-(x_0,x)  \\
-D_+(x_0,x)  & D_D(x_0,x) 
 \end{array}\right),
\end{equation}
where $D_F$, $D_D$, and $D_{-/+}$ are the Feynman, Dyson, and negative/positive Wightman two-point functions, respectively \cite{galley-hu:09}.  Using the saddle-point method, the effective action is readily found to be 
 \begin{equation}\label{eq:eff-action-ctp}
 	\Gamma_{\text{CAP}} [z] = \sum_{a} S_{\text{pp}, \, a}^0 - \frac12 \iint d\tau_{a} d\tau'_{b} \, \,  s^A_{a} D_{AB'}^{ab} s^{B'}_{b} + \frac{i}{2} \iint d\tau_a d\tau'_{b} \, \int dv\,  s^A_{a} D_{AC}^{ac} \mathscr{E}^{CD} D_{DB'}^{c b} s^{B'}_b + \cdots ,
\end{equation}
where all CAP indices are summed over. 
Before taking the variation to obtain the equations of motion for the particle, it is useful to introduce average and difference coordinates 
\begin{align}
	z^\a_+ &:= \frac12(z_1^\a + z_2^\a), \no \\
	 z^\a_-  &:= z_1^\a - z_2^\a .
\end{align} 
The benefit of the above coordinates is that, after variation when we impose the equality condition $z_1^\a = z_2^\a$ , $z^\a_-$  vanishes and $z^\a_+$   can immediately be identified as the physical path. The equations of motion governing the physical path $z_+^\a$ are found by varying $\Gamma_{\text{CAP}} [z]$  with respect to $z_-^\a$ by demanding that the physical path is a stationary point  $ \delta \Gamma_{\text{CAP}} [z] / \delta z_- = 0$. The causal equation of motion is found to be
\begin{equation}\label{eq:eom}
0 = \frac{\delta S_{\text{pp} }^0}{\delta z^\a_-} \Bigg\vert_{z_-=0} + \frac12  F^A_{\ \a} \int d \tau' \,  D^{ret}_{AB'} s^{B'} - \frac{i}{2}  F^A_{\a} \iint \, dv \, d\tau' \, D^{ret}_{AC}\mathscr{E}^{CD} D^{ret}_{DB'} s^{B'} + \cdots,
\end{equation}
where the tensor $F^A_{\ \a}(\tau)$  decomposes into gradient (acting on the Green function) and acceleration components  \[	F^A_{\ \a} := A^A_{\a\b} a^\b + B_{\a\b}^A \nabla^\b \] and has the useful property $D_{AB'} F^{A}_{\ \a} u^\a=0$.  As all world line quantities in the above equation are evaluated on the averaged (physical) configuration, we shall omit the ``+'' index from now on.

\section{Quasi-local expansion}
Although we have succeeded in obtaining a causal equation of motion by  extremizing the effective action, the expression is ill-defined because it contains a divergence at the position of the particle coming from the singular nature of the retarded propagator $D^{\ret} \sim \delta(\sigma)$ on the past lightcone.  In this section we present a local expansion around the accelerated world line for an arbitrary source $\mu^A$ that extends the local treatment of the multi-scalar field to fields with scalar, vector, or tensor representations.

\subsection{Local multi-field self-force I: ignoring curvature contributions from $\mathscr{E}^{AB}$}
We first compute the terms in the equation of motion which do not result from curvature contributions coming from insertions of $\mathscr{E}^{AB}$. The procedure we apply here generalizes to the terms with curvature as well, but the discussion is simplified without their presence. The force which persists in the absence of spacetime curvature under the definition
\begin{equation}\label{eq:force-flat-gen}
\mathfrak{f}_\a(\tau) :=  \frac12 F^A_{\ \a } \int d \tau' \,  D^{\ret}_{AB'} s^{B'}, 
\end{equation}  
where $F^A_{\ \a }$ is evaluated at the present time $\tau$ and the retarded propagator is represented in RNC with origin at $z(\tau)$ as a Fourier integral 
\begin{equation}\label{eq:prop-flat}
D^{\ret}_{AB'}(0,x) = \Lambda_{AB'} \, \int_{k,\, \ret} \frac{e^{- i k\cdot x}}{k^2},
\end{equation}
and where $\Lambda_{AB}$ is defined by the relation $\Lambda_{AC} \Box^{-1} \mathscr{F}^{C}_{\ B} = \delta_{AB}$.  An example of the type of force produced by $\mathfrak{f}_\a(\tau)$ is the Abraham\--Lorentz\--Dirac force describing the radiation damping effect on charged particles in flat spacetime.

The local expansion of Eq.~\eqref{eq:force-flat-gen} proceeds in the same way as in the toy-model: we write the Green function $D_{AB}$ and source $s^A$ as a series in powers of Synge's bivector on the geodesic connecting $z$ and $z'$ and then relate these quantities to the quantities defined on the accelerated world line using the expansion of $\sigma^\mu$ in Eq.~\eqref{eq:sigma-expansion}. This leads to an expression with several terms all involving the integral \eqref{eq:I-master}, which we now examine.  Inspecting \eqref{eq:I-master}  with the flat spacetime propagator $D(k) \sim k^{-2}$,  which implies   $m=1$, we see that the integral yields zero in $d=4$ unless the condition $n=q+p+1$ is satisfied. 

We begin by separating the integral expressions according to whether they involve either  $A^A_{\ \a\b} a^\b$ or $B^A_{\ \a\b}\nabla^\b$ constituting $F^A_{\ \a}$.  First, we consider the $A^A_{\ \a\b} a^\b$ contribution.  Investigating several terms in the local expansion, we find that the condition $n=q+p+1$ is  satisfied by one term only.  The reason is that after removing pieces containing odd powers of spatial vectors $k^{{i}}$, which average to zero, the leading terms are $(1- i  (t^3/3!) \, k^{{0}}  \dot{a}_{{ 0}} )(s^B+ t \dot{s}^B + \cdots )$, and since $q=0$,  the only term remaining is $t \dot{s}^B$.  From this we get a contribution of the form $A^A_{\ \a\b} a^\b {\dot s}^B$. Now, as $s^A$ must  yield a reparameterization invariant action, it must be constructed from the four-velocity alone. It follows that $A^A_{\ \a\b} a^\b {\dot s}^B$ is quadratic in the acceleration.  Higher order terms  in $k^\a$, including those with $q \geq 2$ which involve products of sums  containing multiple powers of $t$ for a single increase in the power of $k^{ 0}$, cannot satisfy $n=p+q+1$ and yield zero.   We now move to the $B^A_{\ \a\b}\nabla^\b$ part of $F^A_{\ \a }$. The presence of the covariant derivative introduces an additional positive power of $k^\a$. Power counting the exponents in the expansion $(k^{ 0} u^\b + k^{ i } e^\b_{ i }) 
(1- i (t^2/2!) \,\, k_{{ j}} a^{{ j}} - i ( t^3/3!) \, ( k^{{ 0}}  \dot{a}_{{ 0}} +  k^{{ i}} \dot{a}_i)   +  \cdots )(s^B+ t \dot{s}^B + \cdots )$, we see that three terms can satisfy the non-zero condition $n=q+p+1$ with $q$ even: two proportional to a derivative of the acceleration and the other one proportional to the acceleration squared. 

 Collecting the above results, we find that the self-force in flat spacetime with a general source $s^A$  is 
\begin{equation}\label{eq:force-flat}
\mathfrak{f}_\a(\tau) = \frac{1}{4\pi} \Lambda_{AB} \Bigg(\tfrac12 A^A_{\ \a\b} a^\b {\dot s}^B + B^A_{\ \a\b} \left( \frac12 \frac{D}{d\tau} ( \dot{s}^B u^\b) 
+ \frac16 s^B \dot{a}^\b \right) \Bigg).
\end{equation}
The equation above can be evaluated immediately once $F^A_{\ \a }$ is determined, and so it provides a very efficient route from the variation \eqref{eq:eom} to a finite part of the self-force in non-vacuum theories.  We now move to deriving a similar expression that includes curvature corrections. 

\subsection{Local multi-field self-force II:  curvature contributions involving $\mathscr{E}^{AB}$}
The focus of this section is the local self-force due to the curvature expansion of the equation of motion \eqref{eq:eom} in powers of  $\mathscr{E}^{CD}$ insertions. First, we introduce the quantity 
\begin{equation}
	\mathpzc{f}_\a := - \frac{i}{2}  F^A_{\a} \iint \, dv \, d\tau' \, D^{\ret}_{AC}\mathscr{E}^{CD} D^{\ret}_{DB'} s^{B'} ,
\end{equation}
 which defines the leading-order, curvature-influenced local force. At this order, the curvature operator only contains single powers of the curvature tensors and they appear undifferentiated. 

To proceed, we require a slight extension of the procedure used in the previous section to accommodate the additional coordinate expansion of the bulk curvature operator $  \mathscr{E}^{CD} = \mathscr{E}^{CD}_{\hat{L}} y^{\hat{L}}$, where $y^\a$ is the RNC of  a \emph{bulk} point in the normal neighbourhood of the origin. The introduction of uppercase Latin letters with carets facilitates the use of multi-indices $\hat{L} = \mu_{1}\cdots \mu_{\ell}$.  In order to use the master integral expression \eqref{eq:I-master} we must recast $\mathpzc{f}_\a$ into the appropriate form. To do this, we need to perform the volume integral. The volume integral in RNC centred at $z(\tau)$ has the form $ \int d^d y \,  y^{\hat L} e^{-i k \cdot  y } e^{i p \cdot (y-x)}$, where, as before, $x^\a$ is the RNC of the ``past'' point $z(\tau')$.   Rewriting $y^{ \HT{L}} e^{-i k \cdot y} $ as $ i^{ \ell} \tfrac{d^\ell }{d k_{\HT{L}}} e^{i k \cdot y}$ and doing $\ell$ integrations by parts reduces the volume integral to a delta function $\delta(p-k)$, leaving a single Fourier integral. Transforming back to global world line coordinates we obtain
\begin{align}\label{eq:force-E-local-1}
	\mathpzc{f}_\a  = &- \frac{i}{2}  F^A_{\a}  \Lambda_{AC} \Lambda_{DB}   \int  d\tau' \, \int_k \frac{1}{k^2} \left[ \mathscr{K}^{CD}_{\HT{L},  \c\delta}  \left( \frac{d^\ell}{ dk_{\HT{L}}} \frac{k^\c}{k^2} \right)   k^\d \right. \\ \no
	 & \left.+ \mathscr{L}^{CD}_{\HT{L},  \c }  \left( \left( \frac{d^\ell}{ dk_{\HT{L}}} \frac{k^\c}{k^2} \right) + k^\gamma    \left( \frac{d^\ell}{ dk_{\HT{L}}} \frac{1}{k^2} \right) \right)  +  \mathscr{M}_{\HT{L} }^{CD}   \frac{  d^\ell}{ dk_{\HT{L}}}\frac{1}{k^2} \right] e^{i k \cdot \sigma} s^{B'}  + \cdots,
\end{align}
where $\mathscr{K}^{CD}_{\HT{L},  \c\delta}$,  $\mathscr{L}^{CD}_{\HT{L},  \c } $, and  $\mathscr{M}_{\HT{L} }^{CD}$  represent coefficients of terms in the bulk curvature operator $\mathscr{E}^{CD}$ with two, one, and zero derivatives, respectively. We therefore refer to these coefficients as the kinetic, angular momentum, and mass tensors.  The tensor  $\mathscr{K}^{CD}_{\HT{L},  \c\delta}$ has the symmetry  $\mathscr{K}^{CD}_{\HT{L},  \c\delta}= \mathscr{K}^{CD}_{\HT{L},  \delta\c}$.  On dimensional grounds,  $\mathscr{K}^{CD}_{\HT{L},  \c\delta}$,  $\mathscr{L}^{CD}_{\HT{L},  \c } $, and   $\mathscr{M}_{\HT{L} }^{CD}$, begin at $\ell=2$, $\ell=1$, $\ell=0$, respectively.   For reference, in the case of the multi-scalar toy-field we had $\mathscr{K}^{CD}_{\a\b,  \c\delta}  = \mathscr{C}^{CD} \left( \frac13 R_{\a\b\c\d} - \frac16 \eta_{\c\d} R_{\a\b} \right)$ and $\mathscr{M}^{CD} = \mathscr{C}^{CD} V''(\Phi)$.  
 
Again, we analyze the self-force expression in terms of the acceleration $A^A_{ \a\b} a^\b$ and gradient $B^{A}_{\ \a\b}\nabla^\b$ pieces that constitute the tensor $F^{A}_{\ \a\b}$. The part involving $A^{A}_{\ \a\b} a^\b$ gives zero due to the additional powers of $1/k$ coming from products of the flat spacetime propagators.  Its leading order piece, which contains the integral $I(2,0,0,0)$ is convergent for large $k$ like $\Lambda^{-1}$. All sub-leading terms in the sum contribute additional $k^{-1}$ powers  and serve to make the result more  convergent.  The gradient component  $B^{A}_{\ \a\b}\nabla^\b$ does contribute local terms to the equation of motion, however.  The first of these terms is derived from the RNC expansion of the metric determinant which enters as the $\ell=2$ component of  the tensor coefficient $\mathscr{K}^{CD}_{\HT{L},  \c\delta}$.  Using the power counting based on our master integral expression, we see that  multipoles higher than  $\ell=2$ converge rapidly in the local expansion. For $\ell=2$, we obtain the integral
\begin{equation*}
	 I_\b^{\ \m\n, \c\delta} =  \int dt \int_k  \frac{k_\b }{k^6} \left( 2 \eta^{ \c (\m } k^{ \n) } +  \eta^{\mn} k^\c - 4 \frac{k^\c k^\m k^\n}{k^2} \right)   k^\delta e^{i k^{ 0} t},
\end{equation*}
which involves combinations of the integrals $I_{ij}(3,0,1,2)=-i/32\pi \delta_{ij}$, $I_{ijkl} (4,0,1,4)=i/192 \pi \delta_{ijkl}$, and $I(4,0,3,2)= - i/192\pi \delta_{ij} $.  Although there are a large number of permutations involved, the expression is simplified somewhat by the orthogonality condition $B^{A}_{\a \b} u^\a=0$ and the need for an even number of spatial $k$ vectors. For $\ell=1$, we find the integral 
\begin{equation*}
 I^{\ \m\n}_{\b} =	\int dt \int_k \frac{k_\b}{k^4}  \left( \eta^{\m\n} - 4 \frac{ k^{\m} k^{\nu}}{ k^2 } \right)  \, e^{i k^0 t},
\end{equation*}
which has non-zero contributions coupling to the vector coefficient $ \mathscr{L}^{CD}_{\HT{L},  \c }$ from the integrals $I_{ij}(3,0,1,2) = -i / 32\pi \delta_{ij} $ and $I(2,0,1,0)=-i/8\pi$.  Lastly, we have the $\ell=0$ expression which is simply the integral  $I(2,0,1,0)$ multiplied by a single four-velocity vector.  All the above integrals may be performed independently of the  representation of the field, but the final expressions are rather lengthy and considerable simplifications occur when a specific choice is made for the theory-dependent quantities $\Lambda^{AB} $ and $B^A_{\ \a\b}$.  Considering their length, we delay their appearance until we specify the field representations in our discussion of the scalarvac and electrovac spacetimes.  

We end this section with a formal expression 
\begin{equation}\label{eq:force-E-local-2}
 \mathpzc{f}_\a(\tau) = -\frac{i}{2} B^{A\ \b}_{\ \a}\Lambda_{AC} \Lambda_{DB} \left[ \mathscr{K}^{CD}_{\m\n,\c\delta} I_\b^{\ \m\n, \c\delta} + \mathscr{L}^{CD}_{\m,\n} I^{\ \m\n}_{\b} + \mathscr{M}^{CD} I_\b \right] s^{B}
 \end{equation}
for the local self-force due to curvature terms in the bulk action as a local function of proper time.  Although Eq.\eqref{eq:force-E-local-2} appears more complicated than its flat spacetime counterpart Eq.~\eqref{eq:force-flat}, as we shall see later when we consider the scalarvac and electrovac space times, the quantities $ \mathscr{K}^{CD}_{\m\n,\c\delta} $, $\mathscr{L}^{CD}_{\m,\n}$, and $\mathscr{M}^{CD}$ are easily obtained from the second variation of the bulk action. Additionally,  most of the contractions involving the Riemann tensor with the integrals  $I_\b^{\ \m\n, \c\delta}$,  $I^{\ \m\n}_{\b}$  and $I_\b$ give zero, leading to roughly the same number of terms as \eqref{eq:force-flat}.  

\section{Scalarvac self-force}\label{sec:sv}
\subsection{Perturbed scalarvac action}
We now move to our first concrete example: the perturbed scalarvac spacetime.  Here we consider a background spacetime with metric $g_{\a\b}$ and a minimally coupled background scalar field $\Phi$ with action given by 
\begin{equation}\label{eq:sv-action}
S_{\text{bulk} } =\frac{1}{2} \int R \, dV - \frac12 \int \left( g^{\a\b} \nabla_\a \Phi \nabla_\b \Phi + 2 V(\Phi) \right) dV .
\end{equation}
In four spacetime dimensions, the background fields evolve dynamically according to the  field equations
\begin{equation}
 R_{\a\b} =  T_{\a\b} -  \tfrac{1}{2}  g_{\a\b} T , 
 \end{equation}
 and 
 \begin{equation}
 \Box \Phi  - V'(\Phi) = 0
\end{equation}
where $V'(\Phi) = dV /d \Phi$.
The stress-energy  tensor of the background is given by 
\begin{align}
  T_{\a\b} &=-\frac{2}{\sqrt{-g}} \frac{\delta S_{\text{matter}}}{\delta g^{\a\b}} \no \\
 &=  \nabla_\a \Phi \nabla_\b \Phi - \frac12 g_{\a\b} \left( g^{\m\n} \nabla_\m \Phi \nabla_\n \Phi + 2 V(\Phi) \right), 
\end{align}
 which implies $\Ricci{_\m\n} =   \nabla_\m \Phi \nabla_\n \Phi +  g_{\mn} V(\Phi) $ for the Ricci tensor.  The background spacetime is perturbed by the presence of a small body, which at significant distances from its centre looks like a point particle with mass $m$, scalar charge $q$, and higher order multipolar couplings which we will ignore here.  The particle moves on a world line $\gamma$ described by parametric relations $z^\m(\tau)$, where $\tau$ is proper time in the background spacetime.    We denote the four-velocity in the background spacetime as $u^\a$ and normalize  it in the background metric such that $g_{\a\b} u^\a u^\b = -1$.   We add to our background action the following point particle interaction term
 \begin{equation}
  S_{\pp} =  - m \int d\tau + q \int d\tau \, \Phi(\tau).
 \end{equation}

  The disturbance introduced by the particle sources a perturbation of the fields defined as the difference between the full non-linear fields, $\varphi$ and $\mathpzc{g}_{\a\b}$, and their background values 
 \begin{align*} 
 f &= \vphi-\Phi \\  \no
 h_{\a\b} &= \mathpzc{g}_{\a\b} - g_{\a\b}.
 \end{align*}
 The two perturbed fields are packaged using our multi-index notation into the doublet $\psi_A := \{f , h_{\a\b} \}$. The variations $\delta  d\tau = -\frac{1}{2} h_{\m\n} u^\m u^\n  d\tau$ and $\delta \Phi =f$ couple to a source doublet  $\mu^A = -\int_\gamma d\tau \delta_4(x,z) s^A(z) $  given by
  \[ s^A := \{q, 1/2 (m-q \Phi) u^\m u^\n \} . \]We shall also require the first two derivatives of the source 
  \[\dot{s}^{A} = \{0, \left(m- q \Phi\right) a^{(\m}u^{\n)} - \frac12 q \dot{\Phi} u^\m u^\n \}, \] and 
  \[\ddot{s}^{A} = \{0, \left(m- q \Phi \right) \left( \dot{a}^{(\m}u^{\n)} + a^\m a^\n \right) - \frac12 q \ddot{\Phi} u^\m u^\n - 2 q \dot{\Phi} a^{(\m} u^{\n)} \} \]
  in the expansion of the world line. 

 To apply the multifield formalism in Sec. \ref{sec:mf}, we  expand the action to second-order in the field perturbations and introduce a gauge. This allows us to compute the propagators associated with the fields $\psi_A$.   After removing terms proportional to the background field equations, we find that the second-order variation of the bulk action reads 
 \begin{align}
	\delta^2 S_{\text{bulk}} = &\! \int  \! \left(  -\tfrac12 \nabla_{\a}f  \nabla^{\a} f - \frac12 V''(\Phi) f^2 +  \Lambda^{\a\b\c\delta} \left( \nabla_\a  \Phi \nabla_\b f  - \frac14 \nabla_\a h  \nabla_\b \right) h_{\c\delta}  \right. \\ \no &+\left. \frac14  \nabla_\c h_{\a\b}\left( \nabla^\b h^{\a\c} - \frac12 \nabla^\c h^{\a\b}\right )  + \frac 14 V(\Phi) \Lambda^{\a\b\c\d} h_{\ab} h_{\cd} - \frac{1}{2} V'(\Phi) f h 		\right)  dV , 
\end{align}
 where 
 \begin{equation*}
 \Lambda^{\a\b\c\d} = \frac{1}{2}\left( g^{\a\c}g^{\b\d} + g^{\a\d}g^{\b\c}-g^{\ab}g^{\c\d} \right)
 \end{equation*}
 is  the identity on the space of symmetric, trace-free, second-rank tensors. Gauge fixing is implemented via the Faddeev-Popov prescription. For this we choose the gauge condition $G_\a =  \frac{1}{2} \left( L_{\a} - 2  f \nabla_\a\Phi \right),$ where $L_{\a} =  \nabla^\b \left( h_{\a\b} - \frac12 g_{\a\b} \pert{}\right) $ is the Lorenz gauge vector. The benefit of choosing this gauge is that it exchanges the derivative coupling between the scalar field and the metric perturbation for an algebraic coupling.  After simplification we find
 \begin{align}
 \delta^2 S_{\text{bulk}} =  -\tfrac12 \int  \Big( f \left( -  \Box  + \mathpzc{M}  \right) f  + \tfrac14 h_{\a\b}  \left(   -\Lambda^{\a\b\c\d}\Box   + M^{\a\b\c\delta} \right)  h_{\c\d} + \,    h^{\a\b}\mathcal{C}_{\ab} \, \,  f \,  \Big)dV,   \no \\
       \end{align}
where 
\begin{align*}
	\mathpzc{M}  &:= V''(\Phi) + 2  \nabla_\a \Phi \nabla^\a \Phi, \\ \no
	M^{\mn\ab} &:= 2  \left(g^{\m\a}  R^{\n\b} -  \Riemann{^\m ^\a ^\n ^\b } \right) - 2  V(\Phi) \Lambda^{\mn\ab}, \\ \no
	\mathcal{C}^{\ab} &:=  2 \nabla^\a \nabla^\b \Phi.  
\end{align*}
After second variational differentiation of the bulk action we find the gauge-fixed field operator, which has the matrix form
\begin{equation}\label{eq:scalarvac-bulk-op}
\tilde O^{AB} =  \left( \begin{array}{cc}
 - \Box + \mathpzc{M} &  \mathcal{C}^{\a\b}  \\
 \mathcal{C}^{\a\b} & \frac{1}{8} \left( M^{(\a\b)(\c\d)} + M^{(\c\d)(\a\b)} \right) - \frac14 \Lambda^{\a\b\c\d} \Box 
 \end{array} \right) .
 \end{equation}
  Notice that in flat spacetime the matrix $O^{AB}$ has the diagonal form
 \[ O^{AB}_{\text{flat}} = \text{diag}\left( -\Box_f, -\frac14 \Lambda^{\a\b\c\d}_f \Box_f \right), \] where $\Box_f$ is the flat spacetime wave operator $\eta^{\mn} \partial_\m \partial_\n$ and $ \Lambda^{\a\b\c\d}_f $ has the index $f$ to indicate that the flat spacetime metric is to be used in the expression for $\Lambda^{\a\b\c\d}$.  To satisfy the equation $ O^{AB}_{\text{flat}} D_{BC} = -\delta^{A}_{ \ C}$, we see that the flat spacetime Green function must equal \[ D_{AB} = \text{diag}\left( \Box_f^{-1}, 4 P_{\a\b\c\d}^f \Box_f^{-1} \right), \] where  $P_{\a\b\c\d}$ is defined such that $P^{\a\b}_{\ \ \m\n} \Lambda^{\m\n}_{\ \c\d} = \frac12 \left( \delta^\a_\c \delta^\b_\d + \delta^\b_\c \delta^\a_\d \right)$.  In $d$ spacetime dimensions $P^{\a\b\c\d}_f$ has the form  $ P_f^{\a\b\c\d} = \frac12 \left( \eta^{\a\c} \eta^{\b\d} + \eta^{\b\c} \eta^{\a\d} - \frac{2}{d-2} \eta^{\a\b}\eta^{\c\d} \right ) $. Note that in four spacetime dimensions $P_{\a\b\c\d}$ and $\Lambda_{\a\b\c\d}$ are equal. Additionally, we have the causality requirement that $\Box^{-1}$ be chosen according to retarded boundary conditions, as dictated by the CAP construction.
 
 \subsection{Local scalarvac self-force I: expansion for vanishing $\mathscr{E}^{AB}$}
 Here we set out to compute the quasi-local expansion of the scalarvac self-force in flat spacetime according to Eq. \eqref{eq:force-flat}.  For this we need to determine the quantity $F^A_{\ \a}$ which results from varying the effective action
 \begin{align} \label{eq:var-eom}
 \frac{\delta}{\delta z^\a} \Big\vert_{\text{CAP} } \int d\tau d\tau' s^A D^{ret}_{AB'} s^{B'} =     2 \int d\tau' \Bigg[ & \left( a_\a + w_\a^{\ \m}\nabla_\m \right) s^A D^{ret}_{AB'} 
   \no \\  &-\left( w^{\m}_{\ \a} \frac{D}{d\tau} + 2 g^{\sigma \mu} a_{(\a} u_{\sigma)} \right) \frac{\delta s^A}{\delta{u^\m}} D^{ret}_{AB'} \Bigg] s^{B'}.
  \end{align}
The  scalar source $s=q$, which has vanishing time derivative, leaves the simple expression
$ q^2 \int d\tau' \left( a_\a + w_{\a}^{\ \m} \nabla_\m \right) D(z,z'), $
from which we readily read off $F_{\a} = 2 q (a_\a + w^{\ \m}_{\a}\nabla_\m)$. Thus, $A_{\a\b} = 2q g_{\a\b} $ and $B_{\a\b} =  2q  w_{\a\b}$. The tensor element of $F^A_{\ \a}$ is complicated by the presence of the time dependent source $(m-q\Phi)u_\a u_\b$. To simplify matters slightly it is useful to introduce a time-dependent mass given by \[ \mathsf{m} := m - q\Phi.  \]  After the variation $\delta s^A /\delta u^\m$,  we find the expression $\frac12 \left( w_\a^{\ \m\n\lambda} \nabla_\lambda - w_\a^{\ \m} a^\n - \frac12 u^\m u^\n a_\a \right)\mathsf{m}(\tau)  D_{\mn\c\d} $ where the derivative acts on both the Green function and the mass function, and \[
w^{\ab\c\d} = \frac12 u^\b u^\c w^{\a\d} - w^{\a \left(\b \right.} u^{\left. \c \right)} u^\d . \] Using the background equation of motion $\mathsf{m} a_\a = q w_{\a}^{\ \b}\nabla_\b \Phi$, we write $w_\a^{ \ \m\n\b} \nabla_\b \mathsf{m}$ as $-\tfrac12 \mathsf{m} u^\m u^\n a_\a + q w_{\a}^{\ (\m} u^{\n)}\dot{\Phi} $ and lump it into the $A^{\m}_{\ \a\b\n}$ piece of $F^{\ \m\n}_{ \a}$  that doesn't include a derivative operator acting on the Green function.   We proceed by separating $F^{\ \m\n}_{\a}$ into pieces which have a derivative on the background scalar field  instead of the original separation involving the acceleration, and those which act on the Green function.    With this, we have $A_{\m}^{\  \a \b \n} = -  q \left(w_\m^{ \ \a\b\n} + w_{\m}^{\ \a} w^{\b\n} + \frac12 u^\a u^\b \right)$  coupling to a derivative acting on the background scalar field,  and $B_\m^{\ \a\b\n} =  \mathsf{m} w_{\m}^{\ \a\b\n}$ coupling to a derivative on $D_{AB}$. 
Substituting these values and the expressions for the derivatives of the sources into Eq.~\eqref{eq:force-flat}, we arrive at the local contribution to the self-force with vanishing $\mathscr{E}^{AB}$
\begin{equation}
  4 \pi \mathfrak{f}_\a =- \frac{11}{3} \mathsf{m}^2  w_\a^{\ \b} \dot{a}_\b + \frac{1}{3} q^2  w_\a^{\ \b} \dot{a}_\b + 7 q \mathsf{m} a_\a \dot{\Phi}, 
\end{equation}
 with the overall factor of $4 \pi$ deviation from the result of \cite{Zimmerman:2014uja} being due to normalization differences in the action.  Notice that we recover the well-known gravitational anti-damping   and scalar damping  terms, $-\tfrac{11}{3} m^2 \dot{a}^\a$ and $\tfrac13 q^2 \dot{a}^\a$ respectively.  We also find a new term, $7 q {\sf m} a_\a \dot{\Phi}$, due to the additional coupling between fields.  To reduce the order of this equation we use the background equation of motion $a_{\a} = \tfrac{q}{\mathsf{m}} w_{\a}^{\ \b} \nabla_\b \Phi$ and its derivative in the orthogonal direction $\dot{a}_\a = \tfrac{q}{\mathsf{m}} w_\a^{\ \b} \left( u^\c \nabla_\c \nabla_\b \Phi + 2 \tfrac{q}{\mathsf{m}}  u^\c \nabla_\c \Phi \nabla_\b \Phi \right)$. After simplification we find 
 \begin{equation}
4 \pi  \mathfrak{f}_\a  = \frac{2}{3}q^2 \left(\frac{q^2}{\mathsf{m}^2} -\frac12\right) w_\a^{\ \b} {\dot \Phi} \nabla_\b \Phi  + \frac13 \mathsf{m} q \left(-11 + \frac{q^2}{\mathsf{m}^2}\right) w_\a^{\ \b} u^\c \nabla_\c \nabla_\b \Phi
 \end{equation}
 for the order-reduced local (flat) self-force in terms of derivatives of the background scalar field.  
 
 \subsection{Local scalarvac self-force II: contributions from $\mathscr{E}^{AB}$}
Here, we calculate the  local self-force resulting from curvature terms in the bulk for the scalarvac theory. These terms are encoded in the curved spacetime propagator through the tensor  $\mathscr{E}^{AB}$. In a general curved spacetime, the global propagator is not exactly solvable but it can be represented as a perturbative expansion in a small region of spacetime known as the convex normal neighbourhood.  One approach, 
 due to Hadamard, is to assume a distributional ansatz with the appropriate causal structure $G_{A B'} (x,x') = U_{AB'} (x,x') \delta(\sigma) + V_{AB'}(x,x') \theta(-\sigma)$, for which the smooth biscalar coefficients $U$ and $V$ are determined recursively using the field equations.  Using the Hadamard form,  one can apply the Detweiler-Whiting regularization method \cite{detweiler-whiting:03}  to obtain the fields responsible  for the self-force. This was the approach used in \cite{Zimmerman:2014uja}.  
  Here we do not employ the Hadamard form, but instead we build the curved spacetime propagator $G_{AB}$ as a series involving the flat spacetime propagator $D_{AB}$ and corrections coming from the curvature terms $\mathscr{E}_{AB}$ in the bulk action just as we did with the toy-model  \eqref{eq:G-toy}.  


To begin, we break up the bulk action into three parts
\begin{equation*}
 \delta^2 S_{\text{bulk}} = S_h + S_f + S_{mix}
\end{equation*}
 consisting of
\begin{align}\label{eq:coor-S2h}
 \!S_h \! &=  - \frac{1}{8} \int \!\! d V \Bigg[  g^{\mn} \Lambda^{\a\b\c\d} \Bigg( \pd_\m h_{\a\b} \pd_\n h_{\c\d} - 4 \Gamma^{\lambda}_{\ \c \n} \pd_\m h_{\a\b} h_{\lambda \d}  + 4  \Gamma^{\lambda}_{\mu\b} \Gamma^{\sigma}_{\nu\gamma} h_{\a\lambda} h_{\delta\sigma} \Bigg)   \! +\! M^{\mn\ab} h_{\m\n} h_{\a\b} \Bigg] , \no \\
 S_f  &= - \frac12 \int dV \Big( g^{\a\b} \pd_\a f \pd_\b f + \left( V''(\Phi) + 2 \pd_\a \Phi \pd^\a \Phi\right ) f^2 \Big), \no \\
 S_{mix} &= -\frac12   \int dV h_{\ab}   \left(2 \pd^\a \pd^\b +  2 g^{\a\c} \Gamma^\b_{\ \c\m} \pd^\m \Phi \right) f,
 \end{align}
  where we have used the coordinate definition of the covariant derivative $ \nabla_\m h_{\a\b} = \pd_\m h_{\a\b} - \Gamma^{\n}_{\ \a\m} h_{\n\b} - \Gamma^{\n}_{\ \b\m}h_{\n\a}$.
In writing the coordinate form above, we have anticipated  the use of Riemann normal coordinates which we now reintroduce. In the local convex neighborhood of $x_0$,  the scale $\mathcal{R}$ at which the metric changes  is much larger than the typical length of the coordinate to a bulk point $y^\a$. RNC therefore facilitate an expansion of background quantities around their local values as a power series in the quantity $ y^\a \pd_\a\sim r/\mathcal{R}$, where $r$ is the magnitude of the geodesic length connecting the two points $x_0$ and $y$. We will refer to this expansion as the near-point expansion and terms with $n$ derivatives of the background fields will enter at $n$-th order in the expansion.   Using the notation of Sec.~\ref{sec:mf},  we introduce a bulk interaction Lagrangian in the local convex normal neighbourhood given by
\begin{align}
   S_{int} &= -\frac{1}{2} \int \psi_A \mathscr{E}^{AB} \psi_B \, dv,  \no \\
      &= -\frac{1}{2} \sum_{\ell=2}^\infty \int d^d y \Bigg[\mathscr{K}^{AB,\a\b}_{\hat{L}} y^{\hat{L}} \pd_\a \psi_A  \pd_{\b} \psi_B + \mathscr{L}^{AB,\a}_{\hat{L}-1} y^{\hat{L}-1} \psi_{A} \pd_\a \psi_B +  \mathscr{M}^{AB}_{\hat{L}-2} y^{\hat{L}-2} \psi_A \psi_B \Bigg], 
\end{align}
where the coefficients $\mathscr{K}^{AB,\a\b}_{\hat{L}}$, $\mathscr{L}^{AB,\a}_{\hat{L}-1}$, and $\mathscr{M}^{AB}_{\hat{L}-2}$ are evaluated at the base point $x_0$. 

 For the first-order self-force, we only need to compute the next-to-leading order curvature corrections to the propagator, and so we drop terms in the near-point expansion of $\mathscr{E}_{AB}$ with more than two derivatives.    Reading off the quantities from equations Eq. \eqref{eq:scalarvac-bulk-op} and Eq. \eqref{eq:coor-S2h}, and dropping terms third order and higher we find 
\begin{equation*}
\mathscr{K}^{AB,\m\n} = \sqrt{-g} \left( \begin{array}{cc}
    g^{\m\n}  & 0   \ \\
 0   &  \frac14  \Lambda^{\a\b\c\d} g^{\m\n}
 \end{array} \right),
 \end{equation*}
\begin{equation*}
\mathscr{L}^{AB,\m} =  \sqrt{-g} \left( \begin{array}{cc}
    0 & 0   \ \\
 0   &  -g^{\m\n} \Lambda^{\a\b\lambda\c} \Gamma^{\delta}_{\ \n\lambda}  \ \end{array} \right),
\end{equation*}
and
\begin{equation*}
\mathscr{M}^{AB} =  \sqrt{-g} \left( \begin{array}{cc}
    V'' + 2\pd_\a \Phi \pd^\a \Phi & 2 \pd_{\a} \pd_{\b} \Phi \ \\ 
     2 \pd_{\a} \pd_{\b} \Phi  
     &  \frac14 M^{\a\b\c\d} 
       \ \end{array} \right).
\end{equation*}
Using the RNC expansion of the background metric 
\begin{align}\label{eq:RNC-g}
 g_{\ab} = \eta_{\ab} - \frac13 R_{\a\c\b\d}\, y^\c y^\d + \cdots , \quad g^{\ab} &= \eta^{\ab} + \frac13 \Riemann{^\a_\c^\b_\d} \,y^\c y^\d + \cdots ,\no \\
\end{align}
and related quantities 
\begin{align}\label{eq:RNC-detg}
 g = 1-\frac{1}{3} R_{\a\b}\, y^\a y^\b + \cdots , \quad  \Gamma^{\a}_{\ \b\c} = -\frac23 \Riemann{^\a _(\b\c)\d} \,y^\d + \cdots,
\end{align}
we find that the kinetic, angular momentum, and mass vertices in RNC read
\begin{subequations}\label{eq:bulk-vertices-scalar}
\begin{equation}
\mathscr{K}^{,\m\n} =   - \frac16 R_{yy}  \eta^{\mn}  + \frac{1}{3} \Riemann{^\m_y^\n_y}  ,
\end{equation}
\begin{align}\label{eq:vertex-K}
 \mathscr{K}^{\a\b\c\d,\m\n}& =   - \frac16 R_{yy} \Lambda^{\a\b\c\d} \eta^{\m\n} + \frac13 R^{\m \ \n}_{\ y \ y}   \Lambda^{\a\b\c\d}  \no \\
    & \quad + \frac13 \eta^{\m\n} \left(  R^{\a \ (\c }_{\ y\  \ y} \eta^{\d) \b} + R^{\b \ (\c }_{\ y \ \ y} \eta^{\d) \a} - \left(R^{\a \ \b}_{\ y \ y}\eta^{\c\d} + R^{\c \ \d}_{\ y \ y} \eta^{\a\b} \right) \right) , 
    \end{align}
\begin{equation}
    \mathscr{L}^{,\m} = 0, 
\end{equation}
\begin{equation}
     \mathscr{L}^{\a\b\c\d,\m} = \frac{8}{3} \eta^{\m\n} \Lambda^{\a\b\tau (\c}  R^{\d)}_{\ \ (\n\tau) y} ,  
\end{equation}
\begin{equation}
     \mathscr{M} = V'' + 2\pd_\a \Phi \pd^\a \Phi,
\end{equation}
\begin{equation}
     \mathscr{M}^{\a\b} = 4 \pd^{(\a} \pd^{\b)} \Phi,
\end{equation}
\begin{equation}
     \mathscr{M}^{\a\b\c\d} = M^{(\a\b)(\c\d)} ,
\end{equation}
 \end{subequations}
where we have introduced the shorthand notation $A_{y} = A_\a y^{\a}$ to indicate contraction with bulk Riemann coordinates.  We now replace powers of the Riemann coordinate with derivatives with respect to $k^\a$ to put the equation in the form of Eq.~\eqref{eq:force-E-local-1}. 

Evaluating Eq. \eqref{eq:force-E-local-2}, we find that the part of $\mathscr{K}_{\a\b\c\d, \m\n} I_\sigma^{\ \mn,\gamma \delta}$ containing the term $ R^{\c \ \d}_{\ \a \ \b} k_{\c}\frac{d^{2}}{d k_\a d k_\b} \frac{k_\d}{k^2}$ vanishes by virtue of the symmetries of the Riemann tensor. We also find that the entire integral $\mathscr{L}_{\a\b\c\d,\m} I_\sigma^{\ \m\n}$ is  canceled tensorially by the second line of  
$\mathscr{K}^{\a\b\c\d,\m\n}$  in \eqref{eq:vertex-K} when coupled  to  $I_\sigma^{\ \mn,\gamma \delta}$.
Inserting the remaining expressions for the bulk vertices Eq.~\eqref{eq:bulk-vertices-scalar} into Eq.~\eqref{eq:force-E-local-2},  eliminating terms by orthogonality, and employing the integral condition $n=p+q+d-2m-1$,  we find non-zero results from the metric determinants in $\mathscr{K}^{,\m\n}$ and $\mathscr{K}^{\a\b\c\d, \m\n}$, the mixing tensor $\mathscr{M}^{\a\b}$, and the Ricci curvature constituents of $\mathscr{M}^{\a\b\c\d}$. In the formulation using the Hadamard form, the part of the form corresponding to the metric determinant pieces of  $\mathscr{K}^{,\m\n}$ and $\mathscr{K}^{\a\b\c\d, \m\n}$ would be the van Vleck determinant in the direct piece of the curved spacetime propagator.  

After performing the integrations we find that the scalar field kinetic term containing the metric determinant gives the local term $\frac{q^2}{6} R_{\a\b}u^\b$, the tensor kinetic term and the tensor mass term give the contributions $\frac{\mathsf{m}^2}{6}R_{\a\b} u^\b$ and $-2\mathsf{m}^2 \nabla_\a \Phi \nabla_\b \Phi u^\b$, respectively. The term in the local force due to the mixing $\mathscr{M}^{\a\b}$  is $2 \mathsf{m}q \nabla_{\a} \nabla_{\b} \Phi$ . 
Combining everything, we find that the local self-force resulting from bulk curvature insertions reads 
\begin{equation}
4 \pi \mathpzc{f_\a} = w_\a^{\ \b} \left( \frac{1}{6}( \mathsf{m}^2 + q^2) R_{\b\c} u^\c -2 \mathsf{m}^2 \,  u^\c  \, \nabla_\b \Phi \nabla_\c \Phi + 2 q \mathsf{m}\, u^\c \nabla_{\b} \nabla_{\c} \Phi \right).
\end{equation}
\subsection{Scalarvac equation of motion} 
Amalgamating the local results just derived with the tail terms coming from the variation of \eqref{eq:var-eom} with the full non-diagonal Green function $G_{AB'}$ we arrive at the covariant equation 
{\small
\begin{align}
\mathsf{m} a^\m =  &q w^{\mn}  \nabla_\n \Phi +  \frac{1}{4\pi} w^{\mn}\Bigg\{ \frac{1}{6} q^2 \left(-1 + 4 \frac{q^2}{m^2} - 11 \frac{m^2}{q^2} \right) u^\a \nabla_\a \Phi \nabla_\n \Phi  
			     	-\frac{1}{3} mq \left(5 - \frac{q^2}{m^2} \right) u^\a \nabla_\a \nabla_\n \Phi \no \Bigg\} \\ &
				+ q^2 w^{\m\n} \nabla_\n \int_{-\infty}^{\tau_-} d\tau'   G^{ret} + \frac12 \mathsf{m} w^{\m\a\b\lambda} \nabla_\lambda \int_{-\infty}^{\tau_-} d\tau' \mathsf{m}' G^{ret}_{\a\b\c'\d'} u^{\c'} u^{\d'} + \frac{q^2}{\mathsf{m}}  w^{\m\n} \nabla_\n \Phi  \int_{-\infty}^{\tau_-} d\tau'   G^{ret} \no \\
	&- \frac12 \mathsf{m} q   \left( w^{\m\a\b\lambda} + \frac12\left( u^\a u^\b w^{\m\lambda} + 2 w^{\a\m} w^{\b\lambda} \right)\right) \nabla_\lambda \Phi
\int_{-\infty}^{\tau_-} d \tau' G^{ret}_{\a\b\c' \d'}u^{\c'} u^{\d'} \no \\
&+ \frac{q^2}{2 {\sf m} } w^{\m\n} \nabla_ \n \int_{-\infty}^{\tau_-} d\tau' { \sf m }' \, \Phi G^{ret}_{ \ \ \cdot \vert \c' \d'} u^{\c'} u^{\d'} + \frac{q {\sf m}^2}{2} w^{\m\n} \int_{-\infty}^{\tau_-} d\tau' \,  \left( \nabla_\n G^{ret}_{\a\b \vert \cdot} u^{\a} u^{\b} - 2 u^\a \nabla_\a G^{ret}_{\n \c \vert \cdot } u^{\c} \right) \no \\
&+ q^2 \dot{\Phi} w^{\mn} u^\c \int_{-\infty}^{\tau_-}  d\tau' \, G^{ret}_{\n \c\vert \cdot} - q \nabla_\lambda \Phi  \int_{-\infty}^{\tau_-} d \tau' \, \left( w^{\mn} w^{\d \lambda} G^{ret}_{\m\d \vert \cdot} + w^{\m\lambda} G^{ret}_{\c\d\vert \cdot} u^\c u^\d \right)
\end{align}}
describing the motion of a point particle under the influence of the first-order coupled self-force in a scalarvac spacetime. Note that we have adopted the ``dot'' notation to distinguish the off-diagonal components of $G_{AB}$ from the diagonal ones. Despite its complicated appearance, several terms carry simple interpretations. The leading order local term is simply the acceleration in the background spacetime. The second local term $\sim q^2 \dot\Phi \nabla_\n \Phi$, coming from the order-reduced acceleration and the Ricci tensor, may be interpreted as a background force on a modified mass $q \dot{\sf m}$.  The third term is a combination of the order-reduced self-acceleration's time derivative and the the mixing tensor $\mathscr{M}_{\a\b}$.  The expression above differs slightly from the one found in \cite{Zimmerman:2014uja} where a small error was made in the Fermi coordinate components of the gradient of the metric perturbation. Specifically, the component $\nabla_a h_{tt}$ in \cite{Zimmerman:2014uja} is missing a term given by $-2 q \dot\Phi a_a$.  We also find that the tail has a much more  complicated structure than in the vacuum situation.  The rich structure of the tail can be seen in the results of \cite{Zimmerman:2014uja}, which use the regular solutions 
\begin{equation} 
\phi[\text{tail}] := 
4\int_{-\infty}^{t^-} (m-q\Phi) G^{\cdot}_{\ |\mu\nu}(x,z) 
u^\mu u^\nu\, d\tau 
+ q \int_{-\infty}^{t^-} G^{\cdot}_{\ |\cdot}(x,z)\, d\tau, 
\label{scalar_phi_tail} 
\end{equation}
and
\begin{equation} 
h_{\alpha\beta}[\text{tail}] := 
4 \int_{-\infty}^{t^-} (m-q\Phi) \Lambda_{\a\b}^{\ \ \c\d}{G}_{\cd|\mu\nu}(x,z) 
u^\mu u^\nu\, d\tau 
+ q \int_{-\infty}^{t^-} \Lambda_{\a\b}^{\ \ \c\d} {G}_{\cd|\cdot}(x,z)\, d\tau,
\label{scalar_h_tail} 
\end{equation}
which both contain off-diagonal contributions from the Green function $G_{AB}$. 
  
 \section{Electrovac self-force}\label{sec:ev}
\subsection{Perturbed electrovac action}
For our second example we consider a  massive, charged  particle moving in an electrovac spacetime. Here we consider a background spacetime with metric $g_{\a\b}$ and a background vector field $A_\a$ with corresponding bulk background  action
\begin{equation}
S_{\text{bulk}} = \frac12 \int R \, dV - \frac12 \int F_{\a\b} F^{\a\b} \, dV, 
\end{equation}
  where $F_{\a\b}$ is the field strength tensor of the electromagnetic field $F_{\a\b} = 2 \nabla_{[ \a} A_{ \b ]} $.  The stress-energy tensor of the bulk matter is given by
  \[ T_{\a\b} =  2 \left( F_{\a}^{\ \m}F_{\b \m} -\frac14 g_{\a\b} F_{\m\n}F^{\m\n} \right) , \] which implies $R=0$. The 
 background field equations  read 
  \begin{align} 
  R_{\a\b}  &= T_{\a\b} \no \\  
  \nabla^\b F_{\a\b} = 0 &, \quad \nabla_{[  \a } F_{ \b\c ]} = 0.
  \end{align}
  As before the background spacetime is perturbed by the presence of the point particle carrying an electric charge $e$ and mass $m$. The particle follows a world line in the background spacetime $\gamma$ parameterized by proper time $\tau$. We denote the particle's four velocity in the background spacetime by $u^\a$ and normalize it in the background metric such that
 $g_{\a\b} u^\a u^\b = -1$ .  The minimally coupled world line action of the point particle is given by 
  \begin{equation}
  S_{\pp}= - m\int d\tau + e \int A_\m u^\m d\tau.
\end{equation} 
The perturbation introduced by the particle sources a perturbation of the fields defined as the difference between the full non-linear fields, $\mathpzc{a}_\a$ and $\mathpzc{g}_{\a\b}$, and their background values 
 \begin{align*} 
 b_\m &= \mathpzc{a}_\m - A_\m  \\  \no
 h_{\a\b} &= \mathpzc{g}_{\a\b} - g_{\a\b}.
 \end{align*}
 The two perturbed fields are packaged using our multi-index notation into the  doublet $\psi_A := \{b_\a , h_{\a\b} \}$. The variations $\delta  d\tau = -\frac{1}{2} h_{\m\n} u^\m u^\n  d\tau$ and $\delta A_\m = b_\m$ couple to a source doublet  $\mu^A = -\int_\gamma d\tau \delta_4(x,z) s^A(z) $ given by
  \[ s^A := \{e u^\a, 1/2 m u^\m u^\n \} . \]We shall also require the first two derivatives of the source 
  \[\dot{s}^{A} = \{e a^\a,  m a^{(\m} u^{\n)} \} \] and 
  \[\ddot{s}^{A} = \{e \dot{a}^\a,   m \left( \dot{a}^{(\m}u^{\n)} + a^\m a^\n \right ) \} \]
  in the expansion of the world line.

  In order to apply the multifield formalism of Sec. \ref{sec:mf}, we must expand the action to second-order in the field perturbations and introduce a gauge. This allows us to compute the propagators associated with the fields $\psi_A$. We find that the second-order variation of the action reads 
 \begin{align}
  \delta^2 S_{\text{bulk}} = & \int \left( 2 \nabla_{[ \a} b_{\b]} \nabla^\b b^\a + h^{\c\d}  \left( 4 F_{\c}^{\ [\a} \delta^{\b ]}_{\ \d} + g_{\c\d}F^{\a\b} \right) \nabla_{\b} b_\a \right. 
  -  \frac18 F_{\a\b} F^{\a\b} \Lambda^{\m\n\c\d} h_{\m\n} h_{\c\d} \no \\ &- \left. \frac12 F^{\a\b} F^{\m\n} h_{\a\m} h_{\b\n} - \frac14 \Lambda^{\a\b\c\d} \nabla_\a h \nabla_\b h_{\c\d} + \frac14 \nabla_\c h_{\a\b} \left( \nabla^\b h^{\a\c} - \frac12 \nabla^\c h^{\a\b} \right) \right) dV.
 \end{align} 
 Unlike the scalarvac scenario, we are unable to find a gauge that eliminates the derivative coupling while maintaining a smooth relation to the Lorenz gauge. We therefore choose the gauge to be the standard Lorenz gauge for both fields, which translates to the gauge doublet $G_A = \{1/2 L_\a, \nabla_\a a^\a\}$ in the multifield language.  After simplification we find that the bulk action reads
 \begin{equation}\label{eq:S2-bulk-em}
  \delta^2 S_{\text{bulk}} =  -\tfrac12 \int  \Big( b_\a \left( -  g^{\a\b} \Box  + \mathfrak{M}^{\a\b}  \right) b_\b  + \tfrac14 h_{\a\b}  \left(   -\Lambda^{\a\b\c\d}\Box   + \mathcal{M}^{\a\b\c\delta}  \right)  h_{\c\d} + \,    h_{\a\b}\mathcal{V}^{\ab\c\d}\nabla_\d b_\c ,  \Big)dV,   \no \\
 \end{equation}
 where 
 \begin{align*}
 \mathfrak{M}^{\a\b} & :=  R^{\a\b} , \\ 
 \mathcal{M}^{\a\b\c\d} &:=  2  \left(g^{\a\c}  R^{\b\d} - \Riemann{ ^\a ^\c ^\b ^\d } \right) +  \frac12 \left(F_{\m\n}F^{\m\n} \Lambda^{\a\b\c\d} + 4 F^{\a\c} F^{\b\d} \right) ,\\
 \mathcal{V}^{\ab\c\d} &:= -  4 F^{\a [\c} g^{\delta] \b} - g^{\a\b} F^{\c\d}   ,
  \end{align*}
   from which we build the gauge-fixed field operator in matrix form
\begin{equation}
\tilde O^{AB} =  \left( \begin{array}{cc}
 -   g^{\a\b} \Box + \mathfrak{M}^{\a\b} & 0 \ \\
   \mathcal{V}^{(\a\b)\c\d} \nabla_\d   & \frac{1}{8} \left( \mathcal{M}^{(\a\b)(\c\d)} + \mathcal{M}^{(\c\d)(\a\b)} \right) - \frac14 \Lambda^{\a\b\c\d} \Box 
 \end{array} \right) .
 \end{equation}

\subsection{Local electrovac self-force I: expansion for vanishing $\mathscr{E}_{AB}$}
As in the scalarvac scenario we first establish the local self-force result in flat spacetime where $\mathscr{E}_{AB} = 0$.   The absence of any coupling between the metric perturbation and the background electromagnetic field on the world line greatly simplifies the local analysis in comparison to the scalarvac scenario.  Using Eq.~\eqref{eq:var-eom} we find that the CAP variation of the effective action yields 
\begin{equation}
 e^2  u^{\beta } \int _{-\infty}^{\infty} d\tau'  \nabla_{ [ \a } D^{ret}_{\b] \gamma' } u^{\gamma'}   + 
	\frac12 m^2 \int_{-\infty}^{\infty}  d\tau'  \left( w_{\a}^{\ \b \c \lambda} \nabla_{\lambda} -  w_\a^{\b} a^\c - \frac12 u^\b u^\c a_\a\right)  D^{ret}_{\b\c \m' \n'}  u^{\m'} u^{\n'} ,
\end{equation}
from which we read off the quantities 
\begin{align}
 A^\m_{\ \a\b} &= 0 , \no \\
 B^\m_{\ \a\b} &=  2 e (u^\m g_{\a\b} - \delta^{\m}_{\ \a} u_\b ), \no \\
 A_{\m}^{ \ \a\b\lambda} & = - m \left( w_\m^{\ \a} g^{\b\lambda} + \frac12 u^\a u^\b \delta_{\m}^{\ \lambda} \right), \no \\
 B_\m^{ \ \a\b\lambda} &= m w_\m^{ \ \a\b\lambda} \no.
\end{align}
Substituting these expressions and the expressions for the time derivatives of the source into Eq.~\eqref{eq:force-flat} gives the result 
\begin{equation}
4 \pi \mathfrak{f}_\a =  \left(  \frac{2}{3} e^2 - \frac{11}{3} m^2   \right)  w_\a^{\ \b} \dot{a}_\b.
\end{equation}
At zeroth-order in $\mathscr{E}_{AB}$, the local self-force for the electrovac spacetime is simply the sum of the well-known gravitational anti-damping force and the Abraham-Lorentz-Dirac force. The additional term that we saw in the scalarvac case is absent here because the mass of the world line  doesn't depend on the background electromagnetic field. 
Using the background equations of motion $ma_\a = e F_{\a \b} u^\b$ and $m \dot{a}_\a = e u^\b u^\c \nabla_\b F_{\a\c} + \frac{e^2}{m} F_{\a\b}F^{\b\c}u_\c$, we  obtain the order-reduced form of the equation of motion
\begin{align}
4 \pi \mathfrak{f}_{\a} = -\frac13 m e \left(11-2\frac{e^2}{m^2}\right) u^\b u^\c \nabla_\c F_{\a\b} - \frac{1}{3} e^2\left(4 - 2\frac{e^2}{m^2} - 5 \frac{m^2}{e^2}\right) F_{\a\b}F^{\b}_{\ \c} u^\c
\end{align}
which is free of runaway solutions.

 \subsection{Local electrovac self-force II: contributions from $\mathscr{E}_{AB}$}
We now compute the local contribution to the equation of motion resulting from bulk curvature corrections to the Green functions. The starting point is the bulk action in coordinate form 
\begin{equation}
 \delta^2 S_{\text{bulk}} = S_h + S_b + S_{mix},
\end{equation}
where 
\begin{align}\label{eq:coord-S2}
 S_h &=  - \frac{1}{8} \int \!\! d V \Bigg[  g^{\mn} \Lambda^{\a\b\c\d} \Bigg( \pd_\m h_{\a\b} \pd_\n h_{\c\d} - 4 \Gamma^{\lambda}_{\ \c \n} h_{\a\b,\m} h_{\lambda \d}  + \orderof{\Gamma^2} \Bigg)     + \mathcal{M}^{\mn\ab} h_{\m\n} h_{\a\b} \Bigg], \no \\
S_b &= -\frac12 \int dV\, \Big( \pd_\a b_\b \pd^\a b^\b \no + 2 g^{\a\b} \left( \pd_\a b_\m \Gamma^\m_{\ \b\d} b^{\d} - \pd_\b b^\m \Gamma^\d_{\ \a\m} b_\d\right) + \orderof{\Gamma^2} \Big), \no \\
S_{mix} &=  -\frac12 \int dV \Big( h_{\a\b} \mathcal{V}^{\a\b\c\d} \pd_\d b_\c + \orderof{\Gamma^2} \Big).
\end{align}
  Inspecting Eqs. \eqref{eq:coord-S2} and dropping terms higher than second-order in background derivatives we find the curvature potentials to be
\begin{equation*}
\mathscr{K}^{AB,\m\n} = \sqrt{-g} \left( \begin{array}{cc}
    g^{\m\n} g^{\a\b} & 0   \ \\
 0   &  \frac14  \Lambda^{\a\b\c\d} g^{\m\n}
 \end{array} \right),
 \end{equation*}
\begin{equation*}
\mathscr{L}^{AB,\m} =  \sqrt{-g} \left( \begin{array}{cc}
    4 \Gamma^{[\a\b]\m} & 0   \ \\
  \mathcal{V}^{\a\b\c\m}   &  -g^{\m\n} \Lambda^{\a\b\lambda\c} \Gamma^{\delta}_{\ \n\lambda}  \ \end{array} \right),
\end{equation*}
and
\begin{equation*}
\mathscr{M}^{AB} =  \sqrt{-g} \left( \begin{array}{cc}
   \mathfrak{M}^{\a\b} & 0  \\  
   0 &  \frac14 \mathcal{M}^{\a\b\c\d}
       \ \end{array} \right).
\end{equation*}
Using the RNC expansions in Eqs.~\eqref{eq:RNC-g} and \eqref{eq:RNC-detg}, the second-order constituents of $\mathscr{E}_{AB}$ take on the values 
\begin{subequations}\label{eq:bulk-vertices}
\begin{equation}
\mathscr{K}^{\a\b,\m\n} =   \frac{1}{3}  \eta^{\ab} \Riemann{^\m_y^\n_y}  + \frac{1}{3} \eta^{\mn} \Riemann{^\a_y^\b_y}   - \frac16 R_{yy}  \eta^{\mn}\eta^{\a\b},
\end{equation}
\begin{align}
 \mathscr{K}^{\a\b\c\d,\m\n}  &=   - \frac16 R_{yy} \Lambda^{\a\b\c\d} \eta^{\m\n} + \frac13 R^{\m \ \n}_{\ y \ y}   \Lambda^{\a\b\c\d}  \no \\ &+ \frac13 \eta^{\m\n} \left(  R^{\a \ (\c }_{\ y\  \ y} \eta^{\d) \b} + R^{\b \ (\c }_{\ y \ \ y} \eta^{\d) \a} - \left(R^{\a \ \b}_{\ y \ y}\eta^{\c\d} + R^{\c \ \d}_{\ y \ y} \eta^{\a\b} \right) \right),
 \end{align}
\begin{equation}
    \mathscr{L}^{\a\b,\m} = - \frac{8}{3} R^{\a(\b\m)}_{\ \ \ \  \ y}, 
\end{equation}
\begin{equation}
    \mathscr{L}^{\a\b\c,\m} = -4 F^{\a[\c} \eta^{\m]\b} - \eta^{\a\b}F^{\c\m} , 
\end{equation}
\begin{equation}
     \mathscr{L}^{\a\b\c\d,\m} = \frac{8}{3} \eta^{\m\n} \Lambda^{\a\b\tau (\c}  R^{\d)}_{\ \ (\n\tau)  y} , 
     \end{equation}
\begin{equation}
\mathscr{M}^{\ab} = R^{\a\b}  ,
\end{equation}
\begin{equation}
     \mathscr{M}^{\a\b\c\d} = \mathcal{M}^{(\a\b)(\c\d)}
 \end{equation}
 \end{subequations}
 where again we adopt the shorthand notation $A_y := A_\a y^\a$ for contractions with Riemann normal coordinates. Again we proceed by replacing powers of the RNCs with derivatives with respect to momentum to write $\mathpzc{f}_\a$ in the form of \eqref{eq:force-E-local-1}. From there, we directly proceed to the expression for   $\mathpzc{f}_\a$ in terms of the integrals in Eq.~\eqref{eq:force-E-local-2}. 
 
 We begin the integrations by considering the $e^2$ terms.  We find that the kinetic tensor contracts with the integral $I_\b^{\ \m\n,\c\d}$ in such a way that it vanishes geometrically due to the symmetries of the Riemann tensor.  The result of integrating the single derivative angular momentum tensor is $\frac{e^2}{4} R_{\a\b}u^\b$, which combines with the algebraic mass term given by  $\frac{7 e^2}{12} R_{\a\b} u^\b$, to give $\frac{5} {6} e^2 R_{\a\b} u^\b =  e^2 \left( \frac13 R_{\a\m} - F_{\a\b}F^{\b}_{\ \m} \right) u^\m$, showing a deviation from the test field electromagnetic local self-force $\frac13 e^2 R_{\a\m}$.  The non-zero   $m^2$ terms are given by a combination of the RNC expansion of the metric determinant in $ \mathscr{K}^{\a\b\c\d,\m\n}$ and the terms in $\mathscr{M}^{\a\b\c\d} $, yielding  $ \frac{1}{6} m^2 R_{\a\b}u^\b$ and $ 2 F_{\a \b}F^{\b}_{\ \c} u^\c$, respectively.  The local contribution from the mixing is simply $ 4 e m F_{\a\b} a^\b$.  Adding the three parts gives the local electrovac self-force from bulk curvature insertions
\begin{equation}
 4 \pi \mathpzc{f}_{\a} = w_\a^{ \ \m} \left( m^2 \left(  \frac16 R_{\m\n}u^\n + 2 F_{\m\n} F^{\n}_{ \ \rho} u^\rho \right) +  \frac{5}{6}  e^2 R_{\m\n} u^\n \ + 4 m e F_{\m\n} a^\n \right).
\end{equation}

\subsection{Electrovac equation of motion} 
Combining the local results with the tail terms coming from the variation in Eq.~\eqref{eq:var-eom}, and writing $R_{\m\n}$ in terms of the field strength tensor,  we arrive at the covariant equation
{\small \begin{align}
m a^\m =  & e F^{\mn} u_{\n} + \frac{1}{4\pi} w^{\mn} \Bigg\{ -\frac13 m e \left(11-2 \frac{e^2}{m^2}\right) u^\lambda u^\sigma \nabla_\sigma F_{\n\lambda} 
 - \frac13 e^2 \left(4 - 2 \frac{e^2}{m^2} - 5 \frac{m^2}{e^2} \right) F_{\n\sigma}F^{\sigma}_{\ \rho}u^{\rho} \Bigg\}\no \\
  &+\frac{m^2}{2}  \left( w^{\m\a\b\lambda} \nabla_\lambda + \frac12 \frac{e}{m} \left( 2 w^{\sigma\beta} w^{\m\a} + u^\a u^\b w^{\m\sigma} \right) F^\rho_{\ \sigma}u_\rho \right)  \int_{-\infty}^{\tau_-}  d \tau'  G^{ret}_{\a\b\c' \d'} u^{\c'} u^{\d'} \no \\
		&+ 2 e^2 u_\lambda \int_{-\infty}^{\tau_-} d\tau' \nabla^{ [ \m } G^{ \lambda ] \,  ret}_{ \ \beta'} u^{\beta'} 
			+ em w^{\m}_{ \ \sigma} u_\lambda u^\b \int_{-\infty}^{\tau_-}  \nabla^{[ \sigma } G^{\lambda ]\, ret}_{\ \beta\vert \a'} u^{\a'} \no \\ 
			&+ em w^{\m}_{ \ \sigma} u_\lambda \int_{-\infty}^{\tau_-} d \tau'  \nabla^{[ \sigma } G^{\lambda ]\, ret}_{\ \vert \a'\b'} u^{\a'} u^{\b'} 
			- em\left(w^{\mn} a^\delta + \frac12 u^\n u^\d a^\m\right) \int_{-\infty}^{\tau_-} d \tau' G^{ret}_{\n\d\vert \b'} u^{\b'}
			 \end{align}}
describing the motion of a charged particle in an electrovac environment.

\section{Conclusion}
In this paper we have derived the first-order equations of motion for a small compact body moving in various non-vacuum spacetimes using techniques from effective field theory.  We began with the simple example of a scalar particle with $N$ charges coupled to an $N$-component scalar field, which allowed us to introduce the effective action procedure. With the toy model we introduced several technical methods such as the local expansions and dimensional regularization that were used throughout the paper to obtain the force at the location of the particle, while keeping within the simplest possible context.  We also introduced our procedure for building the curved  spacetime propagator as a series constructed from the propagator in flat spacetime and vertex corrections.  Using analogies with the toy-scalar model, we then derived a general expression for the self-force for general non-vacuum spacetimes using multifield notation. With this notation, we gave an explicit expression for the local self-force in terms of quantities that can be read off directly from the action along with a set of integrals falling under a single master integral expression.  

We then applied the formalism to derive equations of motion for point particles in scalarvac and electrovac spacetimes as the main results.
 In the scalarvac spacetime, the mass of the world line evolves with the background scalar field, which led to new local world line couplings in the equation of motion relative to the decoupled scenario.   In contrast, the mass of the world line in the electrovac case is conserved and we found that the local self-force at zeroth order in the bulk vertex $\mathscr{E}_{AB}$ was simply the sum of the vacuum and test electrovac results.  The coupling between field perturbations in the bulk does 
 create additional world line terms, however, and the local force in the presence of curvature cannot be written as a simple sum of the decoupled local force terms. In both the scalarvac and electrovac cases the non-local tail terms were written as integrals involving the gravitational and scalar/electromagnetic propagators and additional non-diagonal Green functions.  These non-diagonal Green functions contribute both local terms and tail terms. We obtained their local contributions by building an approximation using the flat-spacetime diagonal Green function with a series of mixing insertions and found that a single mixing insertion was satisfactory to obtain the local force.  
 
 In this paper we limited our focus to the effective action for the world line and left the computation of the regular field for future investigations.  This has the benefit of offering a direct route to the world line without the need of solving the field equations.  Instead, we use easy-to-work-with Green functions defined in flat spacetime and treat curvature corrections perturbatively as interaction vertices.  The drawback is that one must do a separate computation of the effective action with an external  field line to obtain the retarded field. The benefit of using the field equations is that the solutions for the fields are a byproduct of the self-force calculation instead of a separate endeavor.  With the solutions themselves, at least locally, one can then use the Detweiler-Whiting routine to define a singular field from which one can derive regularization parameters for a practical calculation.  The Detweiler-Whiting routine has been applied to an EFT calculation of third-order scalar perturbations, but results for regular metric perturbations are still lacking.  

The equation of motion results presented here and in \cite{Zimmerman:2014uja}, in combination with the analysis in \cite{Linz:2014vja}, promise to open new avenues of research in the self-force program. One future direction is exploring the self-force equations of motion in alternative theories of gravity.  Fundamental scalar fields are predicted to exist in many low energy limits of quantum gravity theories.  Such scalar-tensor theories can be written in the form of \eqref{eq:sv-action} with an appropriate choice of conformal frame. We are currently formulating the self-force problem for general small bodies in these theories using methods present here as well as those in \cite{Zimmerman:2014uja} with eventual plans to explore the weak-field limit.  Additionally, we wish to revisit Hubeny's overcharging scenario. Using the coupled electrovac equation of motion,  one could expand the local analysis conducted by Hubeny, which ignored the tail piece, to include the new local terms generated by the coupling as a preliminary condition for potential overcharging candidates. The candidates could then be followed up with a full numerical self-force calculation via the regular field using the results of   \cite{Zimmerman:2014uja} and \cite{Linz:2014vja}.

\begin{acknowledgments} 
We thank Chad Galley, Eric Poisson, Ira Rothstein, and Misha Smolkin  for helpful comments and discussions.  
\end{acknowledgments}   
\appendix 

\section{Fixing the gauge}
Under infinitesimal gauge transformations the metric perturbation undergoes a change 
\begin{equation*}
h_{\mn} \rightarrow h_{\mn}^{\chi} = h_{\mn} + 2 \chi_{(\m;\n)}.
\end{equation*}
Let us now check whether the modified Lorenz gauge condition for the scalarvac problem $G_\a =  L_{\a} + a \Phi_{;\a} f $  for constant $a$ alters the path integral in any non-trivial way. We consider the gauge transformation
\begin{equation*}
G^\mu(h^\chi,f) = L^{\mu}(h^\chi)+ a f \Phi^{;\mu},
\end{equation*} 
where 
\begin{equation*}
 L^{\mu}(h^\chi) = L^\m(h) + \Box \chi^\m + R^{\m}_{\ \n} \chi^\n.
\end{equation*} 
The functional determinant of $ G^\a(h^\chi)/\delta \chi^\b= \delta^\a_\b \Box + R^\a_{\ \b} $ 
is independent of the dynamical fields $h_{\a\b}$ and $f$ and therefore contributes nothing to correlation functions. 
Now, as the determinant is independent of $h_{\a\b}$ and $f$, we can remove it from the path integral and relabel the integration variable such that 
\begin{equation*}
\int \mathcal{D}  h \mathcal{D}  f e^{i S} = \text{ det } \left( \delta^\a_{\ \b} \Box + R^\a_{\ \b} \right) \int \mathcal{D} \chi \int Dh Df e^{iS[h,f]} \delta(G(h,f)).  
\end{equation*}
Generalizing the gauge condition to read 
\begin{equation*}
G^\m(h,f) \rightarrow G^\m(h,f)- \lambda^\m(x) 
\end{equation*}
we find 
\begin{equation*}
\int \mathcal{D}  h \mathcal{D} f e^{i S} = \text{ det } \left( \delta^\a_{\ \b} \Box + R^\a_{\ \b} \right) \int \mathcal{D}  \chi \int \mathcal{D} h \mathcal{D} f e^{iS[h,f]} \delta(L^\m+a f \Phi^{;\m}-\lambda^\m).  
\end{equation*}
Since the above equation holds for arbitrary functions $\lambda^\m(x)$,
it must also hold for any normalized linear combination of such functions. We choose the normalization $N(\xi)$ with respect to a Gaussian of width $\xi$
\begin{align*}
\int \mathcal{D}  h \mathcal{D} f e^{i S} &= \text{ det } \left( \delta^\a_{\ \b} \Box + R^\a_{\ \b} \right) N(\xi) \int \mathcal{D} \lambda \, e^{-i \int \frac{\lambda_\m \lambda^\m}{2\xi} } \int \mathcal{D}  \chi \int \mathcal{D} h \mathcal{D} f e^{iS[h,f]} \delta(L^\m+a f \Phi^{;\m}-\lambda^\m) \\ \no
&=  \Bigg( \text{ det } \left( \delta^\a_{\ \b} \Box + R^\a_{\ \b} \right) N(\xi)  \int \mathcal{D}  \chi \Bigg)  \int \mathcal{D} h \mathcal{D} f e^{iS[h,f]}  e^{-\tfrac{i}{2\xi} \int (L_\m+a f \Phi_{;\m}) (L^\m+a f \Phi^{;\m})}
\end{align*}
choosing $\xi = \kappa$ and we see that we have recovered the gauge fixing term in \eqref{eq:S-gf}

\bibliography{eft-nonvac}
\end{document}